\documentclass[letterpaper]{article} 
\usepackage[preprint]{aaai2027}  
\usepackage[hyphens]{url}  
\usepackage{graphicx} 
\usepackage{natbib}  
\usepackage{caption} 
\usepackage{algorithm}
\usepackage{algorithmic}

\usepackage{amsmath}
\usepackage{amssymb}
\usepackage{multirow}

\newcommand{\secref}[1]{Section~\ref{#1}}
\newcommand{\figref}[1]{Figure~\ref{#1}}
\newcommand{\tabref}[1]{Table~\ref{#1}}
\newcommand{\myeqref}[1]{Equation~(\ref{#1})}

\usepackage{newfloat}
\usepackage{listings}
\DeclareCaptionStyle{ruled}{labelfont=normalfont,labelsep=colon,strut=off} 
\floatstyle{ruled}
\newfloat{listing}{tb}{lst}{}
\floatname{listing}{Listing}

\usepackage{booktabs}

\title{Gaussian Stippling: Efficient Sorting-Free 3D Gaussian Rendering through Hybrid Sampling and Spatiotemporal Reconstruction}
\author{
    Zijian Huang \equalcontrib,
    Suiliang Mai \equalcontrib,
    Chuankun Zheng,
    Yuan Meng,
    Yuchi Huo\thanks{Corresponding author.}
}
\affiliations{
    State Key Laboratory of CAD\&CG, Zhejiang University, Hangzhou, China
}

\makeatletter
\let\aaai@maketitle\@maketitle
\def\@maketitle{%
  \aaai@maketitle
}
\makeatother

\begin{document}

\maketitle

\begin{abstract}
Conventional 3D Gaussian Splatting (3DGS) requires depth sorting and ordered alpha blending to correctly render overlapping Gaussian primitives. 
Stochastic transparency enables sorting-free rendering by replacing fractional alpha contributions with discrete stochastic visibility samples, but produces substantial spatial and temporal noise at low sample counts. 
We refer to this conversion from continuous Gaussian splats to discrete visibility samples as \textit{Gaussian Stippling}. 
Based on this, we present an efficient order-independent rendering and reconstruction framework that operates directly on unmodified 3DGS assets. 
Our method adaptively integrates primitive-based and fragment-based stippling, exploiting their complementary efficiency regimes to improve rendering throughput. 
To recover high-quality images from sparse stochastic samples, we further introduce a lightweight Gaussian-aware spatiotemporal reconstruction network. 
Using Gaussian-local attributes, the network combines current and reprojected historical observations to suppress stochastic noise. 
A reconstruction network trained across multiple source scenes supports deployment on the evaluated unseen scenes without scene-specific network training or modification of the Gaussian assets. We also demonstrate interactive rendering on mobile hardware.
Under scene-specific training, our 1-spp compact configuration achieves \(2.3\)--\(2.7\times\) the throughput of standard 3DGS at 1080p, while the 16-spp larger configuration achieves higher average PSNR on all three evaluated benchmarks at a greater computational cost.
\end{abstract}


\begin{figure}[t!]
    \centering
    \includegraphics[width=\linewidth]{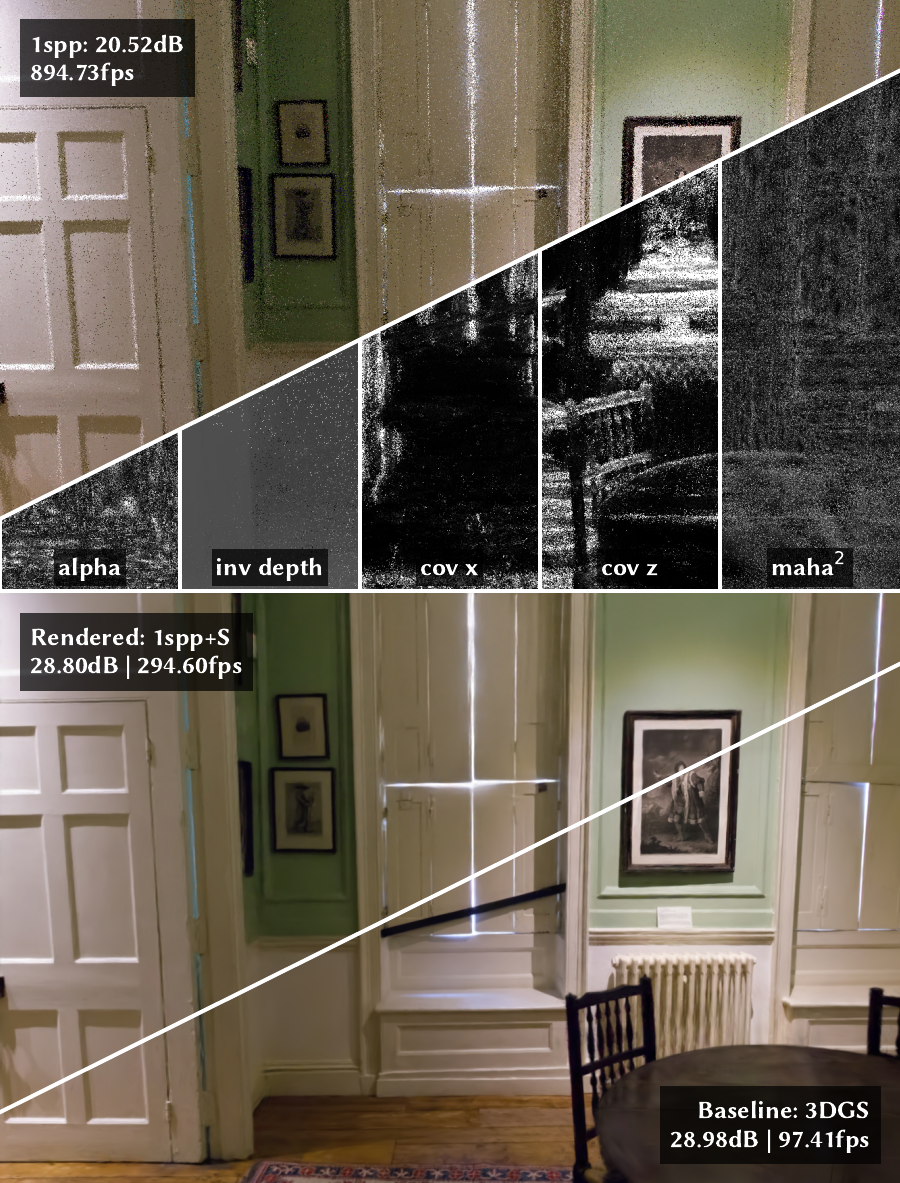}
    \caption{Sorting-free stochastic rendering achieves high throughput but produces substantial spatial noise (top). We augment its observations with Gaussian attributes and reconstruct them using a lightweight spatiotemporal network.}
    \label{fig:teaser}
\end{figure}

\section{Introduction}

\begin{figure*}[t]
    \centering
    \includegraphics[width=\linewidth]{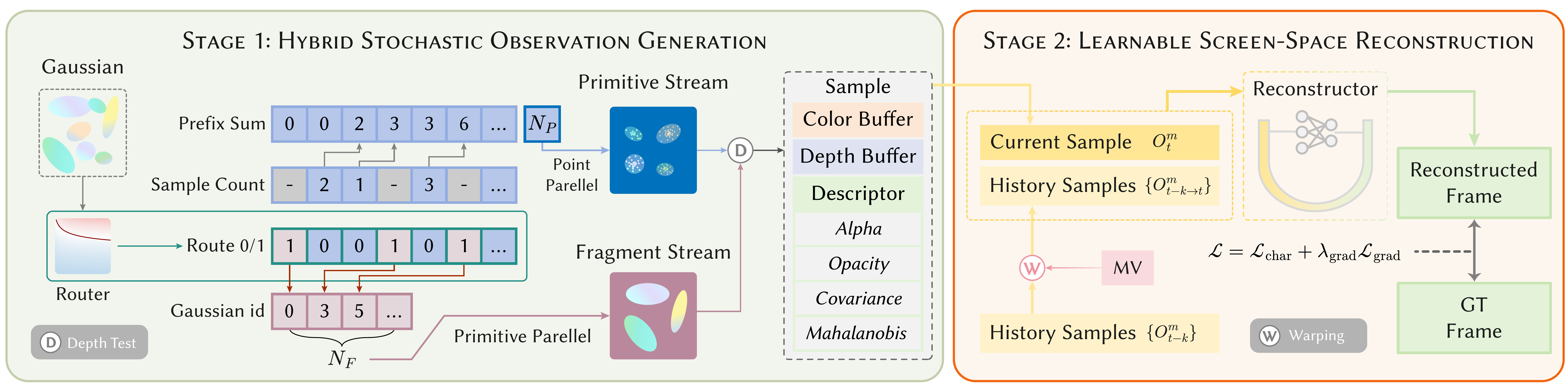}
    \caption{Overview of our proposed Gaussian stippling and reconstruction pipeline. In Stage 1, given a 3DGS asset containing \(N\) Gaussian primitives, a cost-aware routing strategy assigns \(N_F\) primitives to the fragment-based stream and the remaining primitives to the primitive-based stream, which generates \(N_P\) stochastic sample points. A shared depth test then selects accepted stipples together with their Gaussian-local descriptors to generate the current-frame observation. In Stage 2, the reconstruction network aggregates the current observation and forward-reprojected historical observations to produce the reconstructed image.}
    \label{fig:pipeline}
\end{figure*}

3D Gaussian Splatting (3DGS)~\cite{kerbl2023gaussian} enables real-time novel-view synthesis through efficient rasterization of explicit Gaussian primitives. 
However, it relies on depth sorting and ordered compositing of overlapping primitives, which limits its practicality in scenes with large primitive counts~\cite{kerbl2024hierarchical, yang2025virtualized, rijsdijk2026gaussian} and on resource-constrained devices~\cite{du2026mobile, gao2026mobile3dgs3}.
Recent methods alleviate this overhead by reformulating the rendering paradigm~\cite{hou2025sort, muller2025moment, hahlbohm2025efficient, du2026mobile, ye2026depth} and streamlining the primitive representation~\cite{fang2024mini, niedermayr2024compressed}.
Although effective within their respective settings, these approaches are not always directly applicable to existing, unmodified 3DGS assets.

In contrast, stochastic transparency~\cite{enderton2010stochastic} offers a sorting-free mechanism while preserving the underlying Gaussian representation unchanged.
It uses opacity to define stochastic visibility events, allowing Monte Carlo estimation of ordered alpha compositing.
We refer to this transformation as \textit{Gaussian stippling}. 
Conventional Gaussian splatting accumulates continuous Gaussian splats, whereas Gaussian stippling generates discrete, opaque Gaussian visibility samples, termed \textit{stipples}, which compete through a standard depth test. 
This view provides a common description of fragment-based and primitive-based stochastic rendering.

Recent work realizes stochastic transparency efficiently through two strategies, fragment-based stippling~\cite{kheradmand2025stochasticsplats} and primitive-based stippling~\cite{rijsdijk2026gaussian}.
Although both achieve high throughput, the fragment-based cost grows with rasterized coverage, whereas the primitive-based cost grows with the number of generated samples, which depends on opacity and projected footprint. These different cost profiles yield complementary efficiency regimes.  
We exploit this complementarity with a cost-aware hybrid renderer that routes each Gaussian according to its opacity and projected footprint. 
Both branches generate opaque visibility candidates, so their outputs can be combined through a shared depth test.

Efficient sampling alone, however, is insufficient to make low-sample stippling viable for display.
Replacing deterministic compositing with stochastic estimation introduces spatial noise, unstable occlusion boundaries, and temporal flickering. 
Although increasing the sample count can mitigate these artifacts, it also erodes the performance gained from sorting-free rendering.
As illustrated in \figref{fig:teaser}, we address this challenge by using the association between each visible stipple and its originating Gaussian to obtain structured information beyond RGB, including opacity, covariance, and depth.
Moreover, preceding frames can provide additional stochastic samples without increasing the current-frame sampling rate. 
We therefore employ a lightweight spatiotemporal reconstruction network that combines Gaussian-aware observations from the current frame and reprojected historical frames to suppress stochastic noise.

Our experiments evaluate the framework's throughput, reconstruction quality, and cross-scene deployment. 
A spatiotemporal reconstruction network trained across source scenes can be applied to the evaluated unseen scenes without scene-specific network training or modification of the Gaussian assets. We also demonstrate interactive rendering on mobile hardware. 
When scene-specific training is permitted, a higher sampling budget and a larger network provide a quality-oriented operating point at a higher computational cost.


Our main contributions are summarized as follows:
\begin{itemize}
    \item A Gaussian stippling framework for rendering unmodified 3DGS assets, supporting both scene-specific reconstruction and deployment with a fixed cross-scene model, with an implementation on mobile hardware.
    \item An adaptive hybrid stippling renderer that routes Gaussian primitives between primitive-based and fragment-based stipple generation according to their opacity and projected footprint.
    \item A Gaussian-aware spatiotemporal reconstruction method that combines Gaussian-local descriptors from current and reprojected historical observations to suppress stochastic rendering noise.
\end{itemize}

\section{Related Work}
\subsection{Accelerating 3D Gaussian Splatting}

\paragraph{Representation-modified Rendering.} To reduce the cost of depth sorting and alpha compositing in standard 3DGS~\cite{kerbl2023gaussian}, recent methods propose order-independent transparency approximations~\cite{muller2025moment}, depth peeling~\cite{ye2026depth}, or newly optimized Gaussian variants~\cite{hahlbohm2025efficient,du2026mobile,hou2025sort}. Some of these approaches require changes to the Gaussian representation or its optimization procedure. Our renderer instead operates on fixed 3DGS assets. 

\paragraph{Stochastic Visibility Sampling.} Alternatively, render-time acceleration can preserve the assets by altering the rasterization process. Inspired by stochastic transparency~\cite{enderton2010stochastic}, StochasticSplats~\cite{kheradmand2025stochasticsplats} and Gaussian Point Splatting~\cite{rijsdijk2026gaussian} apply Monte Carlo visibility sampling to Gaussians.
Although they eliminate sorting, their low-spp outputs have high variance. Temporal filtering and increased sampling budgets can mitigate this noise. We investigate Gaussian-aware neural reconstruction from low-sample stochastic observations.

\subsection{Neural Reconstruction from Stochastic Samples}

\paragraph{Neural Image-Based Rendering.} 
Synthesizing high-quality images from discrete, unstructured geometric proxies is a core challenge in image-based rendering. 
Neural Point-Based Graphics (NPBG)~\cite{aliev2020neural, rakhimov2022npbg++} and related deferred neural rendering frameworks~\cite{thies2019deferred,ruckert2022adop} have demonstrated that deep convolutional networks (e.g., U-Nets) can effectively reconstruct dense, photorealistic images from rasterized, sparse point features. 
Our setting additionally involves stochastic visibility changes between repeated renders of the same view. 
We therefore adapt the neural deferred rendering paradigm to observations of overlapping, semi-transparent Gaussian primitives.

\paragraph{Reconstruction from Stochastic Samples.} 
Reconstructing expected radiance from high-variance Monte Carlo samples is extensively studied in path tracing~\cite{zwicker2015recent,schied2017spatiotemporal,huo2021survey,bako2017kernel}. 
Deep neural networks are widely used to filter these noisy observations~\cite{vogels2018denoising,fan2021real,balint2023neural}. 
Sample-based reconstruction methods such as SBMC~\cite{gharbi2019sample} motivate the use of information associated with individual stochastic samples. Our method retains Gaussian-local descriptors in each observation; when using multiple observations per frame, it averages these descriptors channel-wise before reconstruction. 
Surface-based auxiliary features, such as depth and albedo, provide useful guidance for reconstruction~\cite{iglesias2020real}, but a single surface description may not capture contributions from overlapping Gaussian splats. 
We therefore extract \textit{Gaussian-local descriptors} that encode the projected footprint, opacity, and depth of sampled primitives to guide reconstruction.

\paragraph{Spatiotemporal Alignment.} 
Given the high variance of 1-spp rendering, temporal reuse can provide additional information for reconstruction~\cite{scherzer2010exploiting, chaitanya2017interactive, bitterli2020spatiotemporal, lin2021fast}. 
Conventional real-time pipelines typically rely on backward reprojection using stable surface motion vectors~\cite{yang2020survey}. 
Under stochastic 3DGS rendering, the visible Gaussian at a pixel may change between repeated renders of the same view, complicating temporal alignment. We use \textit{forward reprojection of historical observation maps}. 
Related neural reconstruction methods learn to combine temporal information~\cite{hofmann2021interactive,zhu2023denoising}. Our network receives current and reprojected historical observations as separate inputs and learns to combine their evidence.

\section{Method}

Given a fixed, trained 3D Gaussian representation \(\mathcal{G}\) and a target camera pose \(C_t\) at frame \(t\), our goal is to reconstruct a high-quality image from a small number of stochastic Gaussian observations without any additional modification or optimization of the original scene representation. As illustrated in \figref{fig:pipeline}, our method consists of two stages. First, a hybrid renderer routes each Gaussian primitive to either a primitive-based sampling stream~\cite{rijsdijk2026gaussian} or a fragment-based sampling stream~\cite{kheradmand2025stochasticsplats} to generate a sparse stipple image of the target view. Second, each accepted sample is augmented with a local descriptor, dubbed \textit{observation}, and historical frames are forward-reprojected into the target view; a lightweight convolutional network then aggregates the resulting spatiotemporal observations to reconstruct the final image.

Specifically, for each frame, we generate \(M\) independent stochastic observation passes and average their descriptors to obtain \(\mathcal{O}_t^M\). We refer to this observation budget as \(M\) spp (samples per pixel). We concatenate the aggregated maps from the current frame and the \(K = 3\) preceding frames into \(X_t\). The reconstruction process is formulated as

\begin{displaymath}
    \hat I_t = D_\theta(X_t),
\end{displaymath}
where \(D_\theta\) denotes the reconstruction network.

\subsection{Hybrid Stochastic Observation Generation}

\paragraph{Stochastic visibility.}

Considering a pixel sample location \(\mathbf{x} \in \mathbb{R}^2\) covered by \(N\) projected Gaussian primitives \(\mathcal{G}_i = \{\boldsymbol{\mu}_i, \boldsymbol{\Sigma}_i, o_i, \mathbf{c}_i, z_i\}\), standard 3D Gaussian splatting~\cite{kerbl2023gaussian} computes front-to-back alpha blending as follows,

\begin{displaymath}
I^{\mathrm{GS}}(\mathbf{x}) = \sum_{i=1}^{N} \alpha_i(\mathbf{x})T_i(\mathbf{x})\mathbf{c}_i,
\end{displaymath}
where \(T_i(\mathbf{x})=\prod_{z_j < z_i}(1 - \alpha_j(\mathbf{x}))\) is the transmittance, \(\alpha_i(\mathbf{x}) = o_i G(\mathbf{x}; \mu_i, \boldsymbol{\Sigma}_i)\) is the alpha value, \(o_i\) is the base opacity, \(\mathbf{c}_i\) is the view-dependent color, and \(z_i\) is the depth.


Stochastic transparency~\cite{enderton2010stochastic}
interprets the effective alpha value as an acceptance probability.
For each Gaussian primitive \(i\) covering pixel \(\mathbf{x}\),
we independently sample a binary acceptance indicator,

\begin{displaymath}
\hat{\alpha}_i(\mathbf{x})
\sim
\operatorname{Bernoulli}\!\left(\alpha_i(\mathbf{x})\right).
\end{displaymath}
When \(\hat{\alpha}_i(\mathbf{x}) = 1\), Gaussian primitive \(i\) is retained as an opaque visibility candidate; otherwise, it is discarded. The closest retained candidate is then selected according to a standard depth test.

Let \(\hat{i}(\mathbf{x})\) and \(\hat{I}^\mathrm{ST}(\mathbf{x})\) denote the selected Gaussian primitive index and the pixel color, respectively. Therefore,

\begin{displaymath}
\begin{aligned}
    \Pr\left[\hat{i}(\mathbf{x}) = i\right] &= \Pr\left[\hat{\alpha}_i(\mathbf{x}) = 1\right] \prod_{z_j<z_i} \Pr\left[\hat{\alpha}_j(\mathbf{x}) = 0\right] \\
                            &= \alpha_i(\mathbf{x})\prod_{z_j<z_i}\left(1-\alpha_j(\mathbf{x})\right) = \alpha_i(\mathbf{x})T_i(\mathbf{x}).
\end{aligned}
\end{displaymath}

Assigning the selected primitive's color to each accepted observation gives, for a zero background contribution,

\begin{displaymath}
\mathbb{E}\!\left[\hat{I}^\mathrm{ST}(\mathbf{x})\right]
= \sum_i \Pr\left[\hat{i}(\mathbf{x}) = i\right]\mathbf{c}_i
= I^{\mathrm{GS}}(\mathbf{x}).
\end{displaymath}
Thus, independent Bernoulli acceptance followed by closest-depth selection is an unbiased estimator of the corresponding ordered alpha-compositing result. Routing preserves this expectation provided that each primitive is assigned to exactly one stream and both streams use the same effective alpha function, support, and independent acceptance events. Pixel discretization and finite-support truncation must be accounted for when applying this idealized argument to an implementation. This unbiasedness statement concerns the stochastic renderer; it does not extend to the learned reconstruction.

\paragraph{Cost-aware routing.}

\begin{figure}[t]
    \centering
    \includegraphics[width=0.8\linewidth]{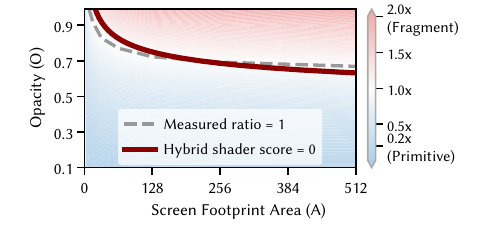}
    \caption{Time-cost ratio of two streams for Layer \(L = 1\). The red solid line is the fitted boundary and the gray dotted line is the measured boundary.}
    \label{fig:score}
\end{figure}

For a Gaussian primitive \(\mathcal{G}\), we characterize its screen-space footprint by the area of its one-standard-deviation ellipse,

\begin{displaymath}
A = \pi\sqrt{\left|\boldsymbol{\Sigma}\right|},
\end{displaymath}
where \(\boldsymbol{\Sigma}\) is its projected 2D covariance.
The fragment-based stream benefits from hardware acceleration to render Gaussian primitives with large projected footprints, but its cost increases with the number of rasterized fragments due to quad overdraw. In contrast, the primitive-based stream avoids fragment rasterization and is more efficient for Gaussian primitives with tiny projected footprints, but its cost grows with the number of generated samples, which depends on both \(A\) and \(o\).
We therefore profile their relative time cost \(\ln(t_{\mathrm{primitive}} / t_{\mathrm{fragment}})\) under different \(o\) and \(A\), which serves as the criterion for determining the faster stream for each Gaussian primitive. To approximate this criterion, we fit the following regression function:

\begin{equation}
S(o, A) = \beta_0 + \beta_1 \log_2 A + \beta_2 o + \beta_3 o \log_2 A,
\label{eq:routing_score}
\end{equation}
where \(\beta_0\), \(\beta_1\), \(\beta_2\) and \(\beta_3\) are the coefficients.
We route Gaussian \(\mathcal{G}\) to the fragment-based stream when \(S(o, A) > 0\), and to the primitive-based stream otherwise.
Details of the profiling procedure and fitting results are presented in \secref{sec:routing_calibration}.

\paragraph{Primitive-based Sampling Stream.}

For Gaussian primitives \(\mathcal{G}\) routed to the primitive-based sampling stream, we follow the opacity-corrected Poisson point process proposed in Gaussian Point Splatting~\cite{rijsdijk2026gaussian}. For each sampling process, the number of generated sample points is drawn from

\begin{displaymath}
n\sim\operatorname{Poisson}(\lambda),
\qquad
\lambda = 2A\operatorname{Li}_2(o),
\end{displaymath}
where \(\operatorname{Li}_2\) denotes the dilogarithm function.

For each sample, we draw a screen-space location \(\mathbf{x}\) from an opacity-corrected spatial distribution using the corrected Box--Muller transform:

\begin{displaymath}
\begin{aligned}
    & u_0, u_1 \sim \operatorname{Uniform}(0, 1),\; r = \sqrt{-2\ln\frac{\operatorname{Li}_2^{-1}(u_0 \operatorname{Li}_2(o))}{o}} \\
    \mathbf{u} &= (r\cdot \cos(2\pi u_1), r\cdot \sin(2\pi u_1))^\top, \qquad \mathbf{x} = \boldsymbol{\mu} + L\mathbf{u},
\end{aligned}
\end{displaymath}
where \(LL^\top=\boldsymbol{\Sigma}\) is the Cholesky decomposition of the covariance matrix.

\paragraph{Fragment-based Sampling Stream}




Following the fragment-based construction of StochasticSplats~\cite{kheradmand2025stochasticsplats}, each Gaussian primitive \(\mathcal{G}\) routed to this stream is rasterized over a finite footprint \(\Omega\). Let \(\tau_F\) denote the fragment alpha cutoff. The effective acceptance probability is

\begin{displaymath}
    \alpha_F^{\mathrm{eff}}(\mathbf{x})
    = \alpha(\mathbf{x})\,
      \mathbf{1}_{\{\mathbf{x}\in\Omega\}}\,
      \mathbf{1}_{\{\alpha(\mathbf{x})>\tau_F\}},
\end{displaymath}
where \(\alpha(\mathbf{x})=oG(\mathbf{x};\boldsymbol{\mu},\boldsymbol{\Sigma})\) is the untruncated Gaussian alpha. Thus, the rasterized footprint and alpha cutoff are part of the effective alpha function used in the visibility argument. Exact cross-stream equivalence requires the primitive-based stream to have the same effective acceptance function, including its support; the conditional argument above does not establish this property for differently truncated implementations.


\paragraph{Cross-stream visibility merging.}

Let \(\mathcal{A}_P(\mathbf{x})\) and \(\mathcal{A}_F(\mathbf{x})\) denote the accepted primitive-based and fragment-based stippling candidates at pixel location \(\mathbf{x}\), respectively.
The hybrid renderer selects that with the smallest depth among all accepted candidates, and stipples its color to pixel \(\mathbf{x}\), denoted by

\begin{displaymath}
\hat{i}(\mathbf{x}) = \underset{p \in \mathcal{A}_P(\mathbf{x})\cup\mathcal{A}_F(\mathbf{x})}{\arg\min} \; z_p,
\qquad
\mathbf{c}(\mathbf{x}) \triangleq \hat{I}^{\mathrm{ST}}(\mathbf{x}) = \mathbf{c}_{\hat{i}(\mathbf{x})}.
\end{displaymath}



\begin{figure}[t]
    \centering
    \includegraphics[width=\linewidth]{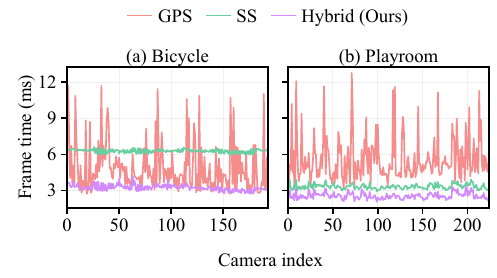}
    \caption{Per-view latency. Ours has the lowest latency among the compared methods on the evaluated views and reduces the viewpoint-dependent variation observed with Gaussian Point Splatting.}
    \label{fig:speed_trace}
\end{figure}

\subsection{Spatiotemporal Gaussian Observations}

Suppose \(\mathcal{G} = \{\boldsymbol{\mu}, \boldsymbol{\Sigma}, o, \mathbf{c}, z\}\) is the final selected Gaussian primitive at pixel location \(\mathbf{x}\) in frame \(t\) for one observation. We associate it with a 10-channel descriptor

\begin{displaymath}
    s_t(\mathbf{x}) = \left[\mathbf{c}(\mathbf{x}), \alpha(\mathbf{x}), \delta^2(\mathbf{x}), o, \mathbf{q}, z^{-1}\right],
\end{displaymath}
where \(\delta^2(\mathbf{x})=(\mathbf{x}-\boldsymbol{\mu})^\top
\boldsymbol{\Sigma}^{-1}(\mathbf{x}-\boldsymbol{\mu})\) is the squared Mahalanobis distance to the projected Gaussian center \(\boldsymbol{\mu}\), \(\mathbf{q}=(\boldsymbol{\Sigma}_{0,0},\boldsymbol{\Sigma}_{0,1},\boldsymbol{\Sigma}_{1,1})^\top\) contains the independent coefficients of the projected 2D covariance matrix \(\boldsymbol{\Sigma}\), and \(z^{-1}\) denotes the inverse depth.
In particular, opacity and Mahalanobis distance help distinguish sampling variations from Gaussian footprint boundaries, while covariance and depth provide anisotropy and visibility cues that are not available from noisy color observations alone.

For a fixed camera pose \(C_t\), we independently sample \(M\) observations and average their descriptors channel-wise to obtain a fixed-dimensional observation map

\begin{displaymath}
    \mathcal{O}_t^M(\mathbf{x}) = \frac{1}{M}\sum_{m=1}^{M}s_t^{(m)}(\mathbf{x}).
\end{displaymath}


Using the current frame and \(K = 3\) historical frames through forward warping~\cite{lin2021fast}, we obtain the network input at pixel \(\mathbf{x}\),

\begin{displaymath}
    X_t(\mathbf{x}) = \operatorname{concat}\left(\tilde{\mathcal{O}}_{t - k \rightarrow t}^M(\mathbf{x})\right)_{k = 0}^{K},
\end{displaymath}
where \(\tilde{\mathcal{O}}_{\tau\rightarrow t}^M\) denotes the observation map from frame \(\tau\) after forward reprojection to frame \(t\), where \(\tilde{\mathcal{O}}_{t\rightarrow t}^M \triangleq \mathcal{O}_t^M\).




\subsection{Neural Reconstruction}



We instantiate the fully convolutional reconstruction network using a pixel-wise MLP bypass with a U-Net~\cite{ronneberger2015u}.
The reconstruction process can be formulated as

\begin{displaymath}
\begin{aligned}
F_{t}^{\mathrm{local}}
    =P(X_t),\qquad F_{t}^{\mathrm{context}} =U_{\theta}(X_t),\\
\hat I_t
    =H_{\theta}\!\left(
      \operatorname{concat}\left(
      F_{t}^{\mathrm{local}},
      F_{t}^{\mathrm{context}}
      \right)\right),
\end{aligned}
\end{displaymath}
where \(P\) is the pixel-wise MLP branch, \(U_{\theta}\) is the multi-scale contextual branch, and \(H_{\theta}\) fuses the two branches to regress the final color.
For the scene-specific ground-truth-supervised configuration, we train the reconstruction network at original dataset views where ground-truth images \(I_t^{\mathrm{GT}}\) are available.
We combine a Charbonnier reconstruction penalty~\cite{charbonnier1997deterministic} with a first-order gradient consistency term, following common practice in image restoration~\cite{zamir2021multi}:

\begin{align}
    \mathcal{L}
    &= \mathcal{L}_{\mathrm{char}}
       + \lambda_{\mathrm{grad}}\mathcal{L}_{\mathrm{grad}},\\
    \mathcal{L}_{\mathrm{char}}
    &= \frac{1}{N}\sum_p
       \sqrt{\|\hat I_t(p)-I_t^{\mathrm{GT}}(p)\|_2^2+\epsilon^2},\\
    \mathcal{L}_{\mathrm{grad}}
    &= \frac{1}{N}\sum_p\sum_{d\in\{x,y\}}
       \|\nabla_d\hat I_t(p)-\nabla_d I_t^{\mathrm{GT}}(p)\|_1,
\end{align}
where \(N\) denotes the number of pixels and \(\epsilon\) is a small constant.
The Charbonnier term provides a robust reconstruction objective, while the gradient term encourages the preservation of local structures and image boundaries.


\section{Experiments}


\begin{table*}[t]
\centering
\small
\setlength{\tabcolsep}{1.5pt}
\begin{tabular}{l|cccc|cccc|cccc}
\hline
\multirow{2}{*}{Method}
& \multicolumn{4}{c}{Mip-NeRF360}
& \multicolumn{4}{c}{Tanks\&Temples}
& \multicolumn{4}{c}{Deep Blending}  \\
\cline{2-13}
& PSNR$\uparrow$ & SSIM$\uparrow$ & LPIPS$\downarrow$ & FPS$\uparrow$ & PSNR$\uparrow$ & SSIM$\uparrow$ & LPIPS$\downarrow$ & FPS$\uparrow$ & PSNR$\uparrow$ & SSIM$\uparrow$ & LPIPS$\downarrow$ & FPS$\uparrow$ \\
\hline
StochasticSplats & \textbf{17.29} & 0.2427 & 0.8942 & \underline{664.95} & \textbf{13.30} & 0.1754 & 0.9961 & \underline{843.08} & \textbf{21.61} & 0.3562 & 0.8063 & \underline{556.03} \\
Gaussian Point Splatting & \underline{17.28} & 0.2413 & 0.8947 & 620.56 & \textbf{13.30} & 0.1749 & 0.9961 & 440.99 & \underline{21.60} & 0.3559 & 0.8065 & 438.95 \\
Ours (1-spp) & \underline{17.28} & 0.2407 & 0.8947 & \textbf{880.07} & \underline{13.29} & 0.1738 & 0.9965 & \textbf{1001.78} & \textbf{21.61} & 0.3559 & 0.8065 & \textbf{1024.15} \\
\hline
3DGS & \underline{28.90} & 0.8708 & 0.1856 & 122.18 & \underline{23.39} & 0.8421 & 0.1837 & 113.19 & 29.52 & 0.9038 & 0.2459 & 111.18  \\
MobileGS\(^\dagger\) & 28.20 & 0.8570 & 0.2103 & 234.97 & -\(^*\) & -\(^*\) & -\(^*\) & -\(^*\)  & \textbf{30.07} & 0.9109 & 0.2486 & 292.26 \\
Mini. + Ours (1-spp + S) & 26.92 & 0.8055 & 0.2607 & \textbf{345.31} & 22.31 & 0.7723 & 0.2830 & \textbf{343.51} & 29.48 & 0.8983 & 0.2703 & \textbf{344.11} \\
Ours (1-spp + S) & 27.72 & 0.8120 & 0.1949 & \underline{283.17} & 22.72 & 0.7780 & 0.2075 & \underline{305.32} & 29.33 & 0.8952 & 0.1895 & \underline{301.10}  \\
Ours (4-spp + S) & 28.23 & 0.8344 & 0.2321 & 159.59 & 22.98 & 0.7904 & 0.2656 & 173.77 & 29.48 &0.8999 & 0.2591 & 160.01  \\
Ours (16-spp + L) & \textbf{29.08} & 0.8606 & 0.1561 & 40.43 & \textbf{23.68} & 0.8252 & 0.1570 & 40.36 & \underline{29.70} & 0.9065 & 0.1735 & 38.05 \\
\hline

\end{tabular}
\caption{Quality against ground truth at the training resolution; throughput at 1080p. The first three rows report renderer-only FPS; Ours-S/L rows include rendering and reconstruction. Best/second-best PSNR and FPS within each block are bold/underlined; ties share rank.
\(^*\) MobileGS optimization produced NaNs on Tanks\&Temples.
\(^\dagger\) MobileGS FPS includes its view-dependent MLP (TensorRT) and rasterization; its released timing code omits the MLP.
}
\label{tab:main_results}
\end{table*}

\subsection{Experimental Setup}

\paragraph{Datasets.}

We evaluate our rendering and reconstruction performance on 11 scenes from Mip-NeRF 360~\cite{barron2022mip}, Tanks and Temples~\cite{knapitsch2017tanks}, and Deep Blending~\cite{hedman2018deep}.
We exclude the `Flower' and `Treehill' scenes from Mip-NeRF 360~\cite{barron2022mip}.
Unless otherwise stated, we train a separate reconstruction network for each scene using its ground-truth training images and evaluate it on held-out views of the same scene; the Gaussian representation remains fixed.
We insert at least three (\(\ge K\)) intermediate camera views between consecutive dataset views, constraining the adjacent camera-center displacement to \(0.05\) reconstructed scene units and the relative rotation to \(3^\circ\).
Ground-truth supervision is applied only to training views; interpolated views are rendered from the fixed Gaussian model and used solely as network inputs.

\paragraph{Baselines.}



We compare our hybrid rendering pipeline and reconstruction network with related rendering methods. StochasticSplats (SS)~\cite{kheradmand2025stochasticsplats} and Gaussian Point Splatting (GPS)~\cite{rijsdijk2026gaussian} provide the fragment-based and primitive-based sampling strategies used by our renderer, respectively. 
MobileGS~\cite{du2026mobile} also eliminates sorting through order-independent transparency (OIT), but requires additional training of the Gaussian representation. 
We also evaluate our pipeline on Gaussian assets produced by MiniSplatting~\cite{fang2024mini}.


\begin{figure}[t!]
    \centering
    \includegraphics[width=\linewidth]{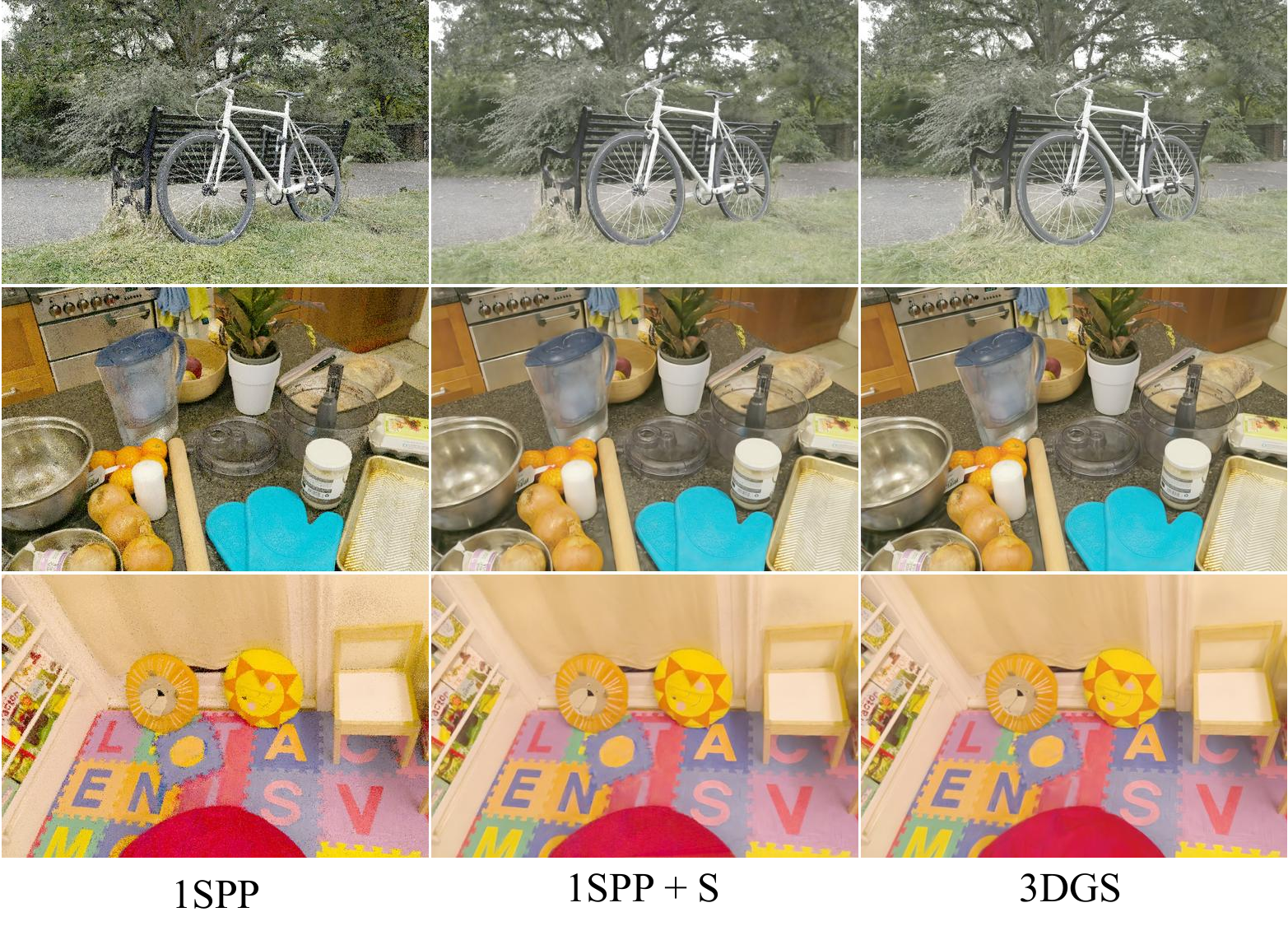}
    \caption{Qualitative results of 1-spp \textit{Ours-S}. A compact reconstruction network suppresses stochastic noise using Gaussian-aware observations from the current and historical frames.}
    \label{fig:qualitativeS}
\end{figure}

\begin{figure}[t!]
    \centering
    \includegraphics[width=\linewidth]{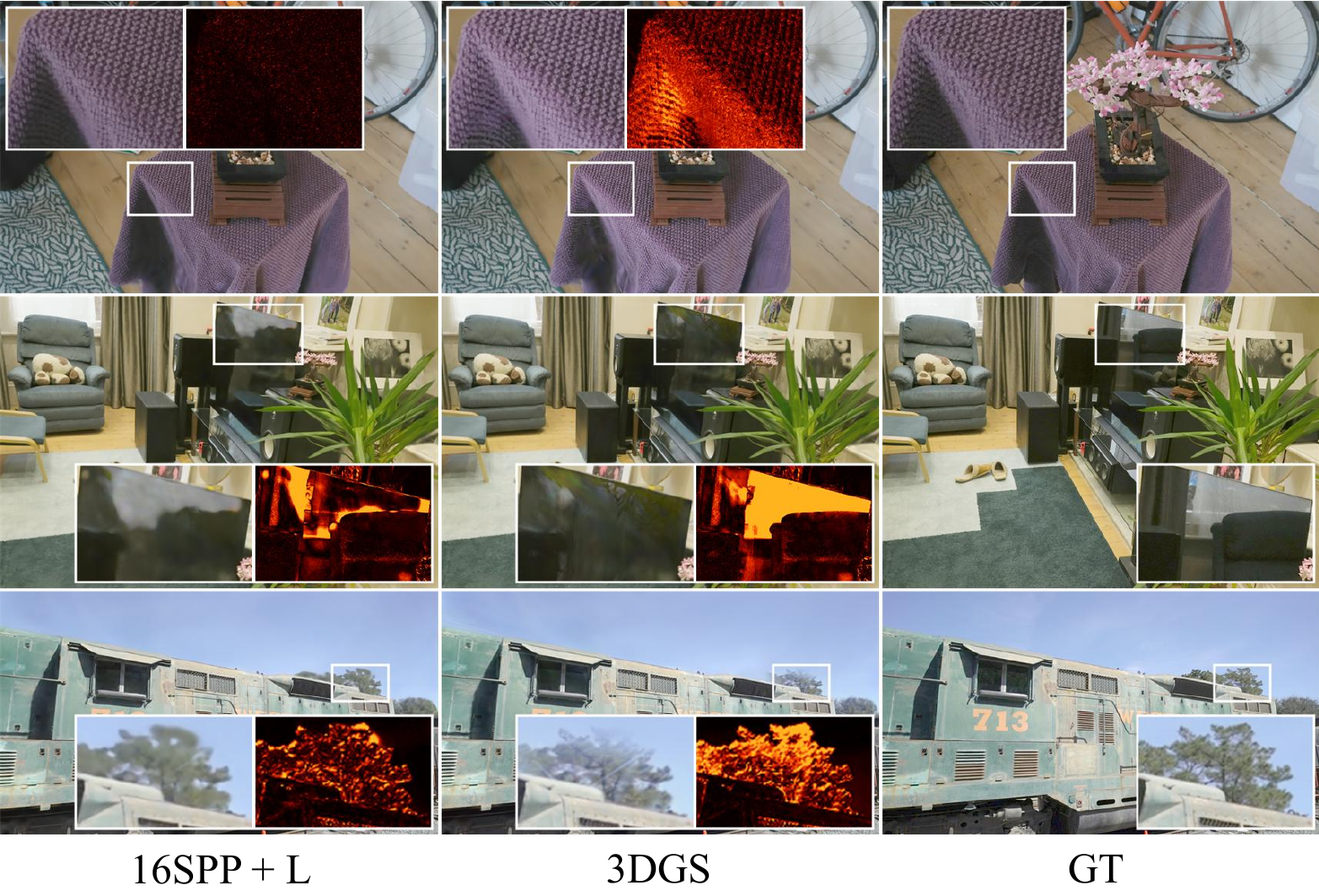}
    \caption{Qualitative results of 16-spp \textit{Ours-L}, illustrating image reconstruction with a higher observation budget and a larger network.}
    \label{fig:qualitativeL}
\end{figure}

\paragraph{Implementation and Training Details.}

We implement reconstruction training in PyTorch and the desktop rendering and inference pipeline using Vulkan and GLSL, with TensorRT~\cite{tensorrt} as the neural inference backend.
Training and desktop inference use GPU acceleration; mobile deployment is described below.
The scene-specific network is optimized for 2,000 epochs with AdamW~\cite{loshchilov2017decoupled}.
In each epoch, seven random \(256 \times 256\) crops are sampled from every training image.
After a 100-epoch linear warmup, the learning rate follows a cosine decay schedule from \(1 \times 10^{-4}\) to \(5 \times 10^{-6}\).
Our compact \textit{Ours-S} and larger \textit{Ours-L} variants both use U-Net-style encoder--decoder architectures but differ in depth, channel widths, and output activation. Their layer configurations, parameter counts, and model sizes are provided in the appendix.

\subsection{Hybrid Stochastic Rendering}
\label{sec:routing_calibration}


\paragraph{Renderer throughput.}

The first three rows of \tabref{tab:main_results} report renderer-only throughput at 1080p. Our hybrid renderer provides a \(7.2\)--\(9.2 \times\) speedup over vanilla 3DGS, while also outperforming both single-stream sampling baselines. \figref{fig:speed_trace} shows lower view-dependent latency fluctuations than the primitive-based stream on the evaluated trajectory.

\paragraph{Routing Criterion Estimation.}

We fit the routing rule using a synthetic benchmark.
Under a fixed camera view, each configuration contains $L$ aligned layers, with a regular \(1000 \times 1000\) grid of Gaussians per layer and \(10^6 L\) Gaussians in total.
All Gaussians share an opacity \(\alpha \in \{0.1,0.2,0.4,0.6,0.8,0.9,0.99\}\) and projected footprint
\(A \in \{2,4,8,16,32,64,128,256,512\}\) pixels.
For each configuration, we record the median 1080p core GPU times of the primitive-based and fragment-based streams, \(t_{\mathrm{primitive}}\) and \(t_{\mathrm{fragment}}\).
The coefficients in \myeqref{eq:routing_score} are then estimated via ordinary least squares.
\figref{fig:score} shows the results for \(L = 1\); results for \(L \in [2, 8]\) are provided in the appendix.
For a fixed \(L\), the two streams exhibit substantial runtime discrepancies under extreme combinations of opacity and footprint. 
Across the tested values of \(L\), the relative cost pattern and crossover boundary remain similar, suggesting that inter-layer occlusion has a limited effect on the streams' relative performance in this synthetic benchmark.

\subsection{End-to-End Reconstruction}

\paragraph{Quality and throughput results.}

\tabref{tab:main_results} compares operating points obtained by varying the observation budget and network capacity. Image quality is evaluated at the training resolution, whereas throughput is measured at 1080p.
At 1 spp, \textit{Ours-S} achieves \(283.17\)--\(305.32\) FPS, providing a \(2.3\)--\(2.7 \times\) speedup over 3DGS. 
At a higher computational cost, the 16-spp \textit{Ours-L} configuration achieves higher average PSNR than 3DGS on the three evaluated benchmarks.
In detail, the reconstruction time of \textit{Ours-S} and \textit{Ours-L} is \(1.78\) ms and \(11.31\) ms, respectively.
Since the reconstruction network operates in screen space, its computational workload depends on image resolution and network capacity rather than the number of Gaussian primitives.
\figref{fig:qualitativeS} shows results of \textit{Ours-S} using one stochastic visibility pass per frame, together with reprojected historical observations.
\figref{fig:qualitativeL} shows results of 16-spp \textit{Ours-L}, including examples of reconstructed fine image detail.







\paragraph{Ablation studies.}
\begin{table}[t]
\centering
\setlength{\tabcolsep}{3pt}
\begin{tabular}{lcccc}
\hline
Variant & Bonsai & Counter & Train & Avg. \\
\hline
Full descriptor                    & 30.749 & 28.263 & 21.450 & 26.820 \\
RGB only                           & 30.376 & 28.093 & 21.117 & 26.529 \\
\hline
w/o alpha                          & 30.732 & 28.253 & 21.439 & 26.808 \\
w/o conic                          & 30.653 & 28.197 & 21.373 & 26.741 \\
w/o opacity                        & 30.758 & 28.249 & 21.421 & 26.809 \\
w/o $d_M^2$                        & 30.710 & 28.263 & 21.453 & 26.809 \\
w/o inverse depth                  & 30.738 & 28.260 & 21.436 & 26.811 \\
w/o opacity and $d_M^2$            & 30.748 & 28.242 & 21.434 & 26.808 \\
\hline

w/o history & 29.999 & 27.883 & 21.160 & 26.347 \\
w/ 1 history frame & 30.398 & 28.105 & 21.351 & 26.618 \\

\hline
\end{tabular}

\caption{
Ablation of temporal history and Gaussian-local descriptors using \textit{Ours-S}. We evaluate variants that remove all historical samples, keep only a single history frame, or omit specific descriptor channels from the full input. All values are PSNR evaluated against the standard 3DGS target. 
}
\label{tab:ablation_input}
\end{table}

All ablations use 1-spp \textit{Ours-S} with the same training settings on three representative scenes. \tabref{tab:ablation_input} evaluates temporal history and Gaussian-local descriptors.
Removing historical observations or retaining only RGB lowers average reconstruction PSNR in these ablations.
Individual descriptor removals have smaller and scene-dependent effects. These results support the usefulness of temporal history and the descriptor set as a whole, while providing limited evidence for the independent contribution of each descriptor.

\paragraph{Generalization.}

For cross-scene evaluation, we train one \textit{Ours-S} reconstruction network on the source scenes described above and evaluate it on five scenes from DL3DV. \tabref{tab:dl3dv_psnr} compares its PSNR and throughput with those of 4-spp StochasticSplats and Gaussian Point Splatting. On these target scenes, the reconstructed outputs achieve higher PSNR than the raw stochastic baselines, while remaining below standard 3DGS on average.

\begin{table}[t]
\centering

\setlength{\tabcolsep}{2pt}
\renewcommand{\arraystretch}{1.08}

\begin{tabular}{@{}lcccccc@{}}
\toprule
Method
& Flow.
& House
& Pavil.
& Plant
& Stat.
& Avg. \\
\midrule

3DGS
& \textbf{22.73}
& \textbf{23.10}
& \textbf{21.25}
& \textbf{22.76}
& \textbf{21.62}
& \textbf{22.29} \\

Ours (1-spp)
& 13.47
& 13.45
& 12.67
& 13.92
& 13.27
& 13.36 \\

SS (4-spp)
& 17.80
& 17.86
& 16.86
& 18.23
& 17.39
& 17.63 \\

GPS (4-spp)
& 17.83
& 17.87
& 16.86
& 18.22
& 17.51
& 17.66 \\

Ours (1-spp + S)
& \underline{21.74}
& \underline{22.35}
& \underline{21.01}
& \underline{21.60}
& \underline{20.94}
& \underline{21.53} \\

\bottomrule
\end{tabular}

\caption{PSNR on held-out DL3DV views with respect to photographic ground truth. Avg. is the arithmetic mean over the five scenes. The best and second-best values are shown in bold and underlined, respectively. SS (4-spp), GPS (4-spp), and Ours (1-spp + S) run at 353.32, 282.91, and 353.43 FPS, respectively.}

\label{tab:dl3dv_psnr}

\end{table}

\subsection{Application: Mobile Deployment and Simultaneous Graphics and Compute}


We further implement our method on a mobile platform to evaluate its applicability on resource-constrained devices.
At 540p, our renderer achieves an average throughput of 92.6 FPS, compared with 53.3 FPS and 49.0 FPS for StochasticSplats and Gaussian Point Splatting, respectively.
With the compact reconstruction network, our implementation achieves 73.2 FPS on the evaluated mobile platform using its neural processing unit (NPU) for inference. The 3DGS-GL baseline uses Splatapult~\cite{anthony2023splatapult}. Our interactive implementation additionally applies reprojected temporal accumulation to the reconstructed images, as described in the appendix.

\begin{table}[!ht]
\centering
\begin{tabular}{lcccc}
\hline
Method & Truck & Room & DrJohnson & Avg. \\
\hline
3DGS-GL                & 5.46 & 12.8 & 3.2 & 7.15 \\
SS             & 24.8 & 65.8 & \underline{69.4} & 53.3 \\
GPS            & 41.3 & 52.8 & 53.0 & 49.03 \\
SortfreeGS & 37.2 & 50.3 & 27.2 & 38.2 \\
MobileGS & -- & 32.6 & 28.1 & 30.4 \\
Ours(1-spp)                  & \textbf{78.8} & \textbf{112.7} & \textbf{86.3} & \textbf{92.6} \\

Ours(1-spp + S)         & \underline{64.6} & \underline{86.7} & 68.3 & \underline{73.2} \\
\hline
\end{tabular}
\caption{Mobile 540p throughput (FPS) on vivo x300pro with MediaTek tianji9500, using one scene per benchmark. Our renderer uses Vulkan, and Ours-S runs in FP16 on the NPU. Avg. is the arithmetic mean over available scene results. The best and second-best values are shown in bold and underlined, respectively.}
\label{tab:mobile}
\end{table}

\section{Conclusion}
We presented a Gaussian stippling framework for efficient sorting-free 3D Gaussian rendering. 
By converting continuous Gaussian splats into discrete stochastic visibility samples, our hybrid sampler combines the complementary efficiency regimes of primitive-driven and fragment-driven sampling without modifying the underlying 3DGS assets. 
Gaussian-aware spatiotemporal reconstruction then combines current and historical stochastic evidence to suppress noise in low-sample renderings. 
The framework supports both deployment with a fixed cross-scene reconstruction model and a scene-specific configuration that prioritizes reconstruction quality.

Our main limitation is that reconstruction from sparse stochastic observations can over-smooth high-frequency appearance, particularly around thin structures, sharp visibility transitions, and view-dependent details. 
This behavior may reflect both the limited evidence available at low sampling rates and the reconstruction objective. 
A promising direction is unified end-to-end per-scene training that jointly optimizes the Gaussian representation, sampling strategy, and reconstruction network for the final image objective. 
Such a quality-oriented mode could better preserve fine detail, but would require additional scene-specific optimization beyond the fixed-asset deployment considered here.

\bibliography{aaai2027}

\begin{thebibliography}{46}
\providecommand{\natexlab}[1]{#1}

\bibitem[{Aliev et~al.(2020)Aliev, Sevastopolsky, Kolos, Ulyanov, and
  Lempitsky}]{aliev2020neural}
Aliev, K.-A.; Sevastopolsky, A.; Kolos, M.; Ulyanov, D.; and Lempitsky, V.
  2020.
\newblock Neural point-based graphics.
\newblock In \emph{European conference on computer vision}, 696--712. Springer.

\bibitem[{Bako et~al.(2017)Bako, Vogels, McWilliams, Meyer, Nov{\'a}k, Harvill,
  Sen, Derose, and Rousselle}]{bako2017kernel}
Bako, S.; Vogels, T.; McWilliams, B.; Meyer, M.; Nov{\'a}k, J.; Harvill, A.;
  Sen, P.; Derose, T.; and Rousselle, F. 2017.
\newblock Kernel-predicting convolutional networks for denoising Monte Carlo
  renderings.
\newblock \emph{ACM Trans. Graph.}, 36(4): 97--1.

\bibitem[{Balint et~al.(2023)Balint, Wolski, Myszkowski, Seidel, and
  Mantiuk}]{balint2023neural}
Balint, M.; Wolski, K.; Myszkowski, K.; Seidel, H.-P.; and Mantiuk, R. 2023.
\newblock Neural partitioning pyramids for denoising monte carlo renderings.
\newblock In \emph{ACM SIGGRAPH 2023 conference proceedings}, 1--11.

\bibitem[{Barron et~al.(2022)Barron, Mildenhall, Verbin, Srinivasan, and
  Hedman}]{barron2022mip}
Barron, J.~T.; Mildenhall, B.; Verbin, D.; Srinivasan, P.~P.; and Hedman, P.
  2022.
\newblock Mip-nerf 360: Unbounded anti-aliased neural radiance fields.
\newblock In \emph{Proceedings of the IEEE/CVF conference on computer vision
  and pattern recognition}, 5470--5479.

\bibitem[{Bitterli et~al.(2020)Bitterli, Wyman, Pharr, Shirley, Lefohn, and
  Jarosz}]{bitterli2020spatiotemporal}
Bitterli, B.; Wyman, C.; Pharr, M.; Shirley, P.; Lefohn, A.; and Jarosz, W.
  2020.
\newblock Spatiotemporal reservoir resampling for real-time ray tracing with
  dynamic direct lighting.
\newblock \emph{ACM Transactions on Graphics (TOG)}, 39(4): 148--1.

\bibitem[{Chaitanya et~al.(2017)Chaitanya, Kaplanyan, Schied, Salvi, Lefohn,
  Nowrouzezahrai, and Aila}]{chaitanya2017interactive}
Chaitanya, C. R.~A.; Kaplanyan, A.~S.; Schied, C.; Salvi, M.; Lefohn, A.;
  Nowrouzezahrai, D.; and Aila, T. 2017.
\newblock Interactive reconstruction of Monte Carlo image sequences using a
  recurrent denoising autoencoder.
\newblock \emph{ACM Transactions on Graphics (TOG)}, 36(4): 1--12.

\bibitem[{Charbonnier et~al.(1997)Charbonnier, Blanc-F{\'e}raud, Aubert, and
  Barlaud}]{charbonnier1997deterministic}
Charbonnier, P.; Blanc-F{\'e}raud, L.; Aubert, G.; and Barlaud, M. 1997.
\newblock Deterministic edge-preserving regularization in computed imaging.
\newblock \emph{IEEE Transactions on image processing}, 6(2): 298--311.

\bibitem[{Du(2026)}]{xiaobiaodu2026mobilegs}
Du, X. 2026.
\newblock Mobile-GS Github source code.

\bibitem[{Du et~al.(2026)Du, Wang, Zhan, and Yu}]{du2026mobile}
Du, X.; Wang, Y.; Zhan, K.; and Yu, X. 2026.
\newblock Mobile-GS: Real-time Gaussian splatting for mobile devices.
\newblock 2026.

\bibitem[{Enderton et~al.(2010)Enderton, Sintorn, Shirley, and
  Luebke}]{enderton2010stochastic}
Enderton, E.; Sintorn, E.; Shirley, P.; and Luebke, D. 2010.
\newblock Stochastic transparency.
\newblock In \emph{Proceedings of the 2010 ACM SIGGRAPH symposium on
  Interactive 3D Graphics and Games}, 157--164.

\bibitem[{Fan et~al.(2021)Fan, Wang, Huo, and Bao}]{fan2021real}
Fan, H.; Wang, R.; Huo, Y.; and Bao, H. 2021.
\newblock Real-time Monte Carlo denoising with weight sharing kernel prediction
  network.
\newblock In \emph{Computer Graphics Forum}, volume~40, 15--27. Wiley Online
  Library.

\bibitem[{Fang and Wang(2024)}]{fang2024mini}
Fang, G.; and Wang, B. 2024.
\newblock Mini-splatting: Representing scenes with a constrained number of
  gaussians.
\newblock In \emph{European conference on computer vision}, 165--181. Springer.

\bibitem[{Gao et~al.(2026)Gao, Zhao, Song, Wen, Song, Cai, and
  Liu}]{gao2026mobile3dgs3}
Gao, F.; Zhao, Y.; Song, C.; Wen, J.; Song, Y.; Cai, Y.; and Liu, L. 2026.
\newblock Mobile3DGS3: Accelerate Mobile 3DGS Rendering via Gradient-Aware
  Super-Sampling and Frame Interpolation.
\newblock In \emph{Proceedings of the Special Interest Group on Computer
  Graphics and Interactive Techniques Conference Conference Papers}, 1--11.

\bibitem[{Gharbi et~al.(2019)Gharbi, Li, Aittala, Lehtinen, and
  Durand}]{gharbi2019sample}
Gharbi, M.; Li, T.-M.; Aittala, M.; Lehtinen, J.; and Durand, F. 2019.
\newblock Sample-based Monte Carlo denoising using a kernel-splatting network.
\newblock \emph{ACM Transactions on Graphics (ToG)}, 38(4): 1--12.

\bibitem[{Hahlbohm et~al.(2025)Hahlbohm, Friederichs, Weyrich, Franke, Kappel,
  Castillo, Stamminger, Eisemann, and Magnor}]{hahlbohm2025efficient}
Hahlbohm, F.; Friederichs, F.; Weyrich, T.; Franke, L.; Kappel, M.; Castillo,
  S.; Stamminger, M.; Eisemann, M.; and Magnor, M. 2025.
\newblock Efficient Perspective-Correct 3D Gaussian Splatting Using Hybrid
  Transparency.
\newblock In \emph{Computer Graphics Forum}, volume~44, e70014. Wiley Online
  Library.

\bibitem[{Hedman et~al.(2018)Hedman, Philip, Price, Frahm, Drettakis, and
  Brostow}]{hedman2018deep}
Hedman, P.; Philip, J.; Price, T.; Frahm, J.-M.; Drettakis, G.; and Brostow, G.
  2018.
\newblock Deep blending for free-viewpoint image-based rendering.
\newblock \emph{ACM Transactions on Graphics (ToG)}, 37(6): 1--15.

\bibitem[{Hofmann et~al.(2021)Hofmann, Hasselgren, Clarberg, and
  Munkberg}]{hofmann2021interactive}
Hofmann, N.; Hasselgren, J.; Clarberg, P.; and Munkberg, J. 2021.
\newblock Interactive path tracing and reconstruction of sparse volumes.
\newblock \emph{Proceedings of the ACM on Computer Graphics and Interactive
  Techniques}, 4(1): 1--19.

\bibitem[{Hou et~al.(2025)Hou, Rauwendaal, Li, Le, Farhadzadeh, Porikli, Bourd,
  and Said}]{hou2025sort}
Hou, Q.; Rauwendaal, R.; Li, Z.; Le, H.; Farhadzadeh, F.; Porikli, F.; Bourd,
  A.; and Said, A. 2025.
\newblock Sort-free gaussian splatting via weighted sum rendering.
\newblock In \emph{International Conference on Learning Representations},
  volume 2025, 5655--5671.

\bibitem[{Huo and Yoon(2021)}]{huo2021survey}
Huo, Y.; and Yoon, S.-e. 2021.
\newblock A survey on deep learning-based Monte Carlo denoising.
\newblock \emph{Computational visual media}, 7(2): 169--185.

\bibitem[{Iglesias-Guitian, Mane, and Moon(2020)}]{iglesias2020real}
Iglesias-Guitian, J.~A.; Mane, P.; and Moon, B. 2020.
\newblock Real-time denoising of volumetric path tracing for direct volume
  rendering.
\newblock \emph{IEEE Transactions on Visualization and Computer Graphics},
  28(7): 2734--2747.

\bibitem[{Kerbl et~al.(2023)Kerbl, Kopanas, Leimkuehler, and
  Drettakis}]{kerbl2023gaussian}
Kerbl, B.; Kopanas, G.; Leimkuehler, T.; and Drettakis, G. 2023.
\newblock {3D Gaussian Splatting for Real-Time Radiance Field Rendering}.
\newblock \emph{ACM Transactions on Graphics}, 42(4): 1--14.

\bibitem[{Kerbl et~al.(2024)Kerbl, Meuleman, Kopanas, Wimmer, Lanvin, and
  Drettakis}]{kerbl2024hierarchical}
Kerbl, B.; Meuleman, A.; Kopanas, G.; Wimmer, M.; Lanvin, A.; and Drettakis, G.
  2024.
\newblock A hierarchical 3d gaussian representation for real-time rendering of
  very large datasets.
\newblock \emph{ACM Transactions On Graphics (TOG)}, 43(4): 1--15.

\bibitem[{Kheradmand et~al.(2025)Kheradmand, Vicini, Kopanas, Lagun, Yi,
  Matthews, and Tagliasacchi}]{kheradmand2025stochasticsplats}
Kheradmand, S.; Vicini, D.; Kopanas, G.; Lagun, D.; Yi, K.~M.; Matthews, M.;
  and Tagliasacchi, A. 2025.
\newblock Stochasticsplats: Stochastic rasterization for sorting-free 3d
  gaussian splatting.
\newblock In \emph{Proceedings of the IEEE/CVF International Conference on
  Computer Vision}, 26326--26335.

\bibitem[{Knapitsch et~al.(2017)Knapitsch, Park, Zhou, and
  Koltun}]{knapitsch2017tanks}
Knapitsch, A.; Park, J.; Zhou, Q.-Y.; and Koltun, V. 2017.
\newblock Tanks and temples: Benchmarking large-scale scene reconstruction.
\newblock \emph{ACM Transactions on Graphics (ToG)}, 36(4): 1--13.

\bibitem[{Li(2025)}]{yukeli2025sortfreegs}
Li, Y. 2025.
\newblock Sortfree-GS Github source code.

\bibitem[{Lin, Wyman, and Yuksel(2021)}]{lin2021fast}
Lin, D.; Wyman, C.; and Yuksel, C. 2021.
\newblock Fast volume rendering with spatiotemporal reservoir resampling.
\newblock \emph{ACM Transactions on Graphics (TOG)}, 40(6): 1--18.

\bibitem[{Ling et~al.(2024)Ling, Sheng, Tu, Zhao, Xin, Wan, Yu, Guo, Yu, Lu
  et~al.}]{ling2024dl3dv}
Ling, L.; Sheng, Y.; Tu, Z.; Zhao, W.; Xin, C.; Wan, K.; Yu, L.; Guo, Q.; Yu,
  Z.; Lu, Y.; et~al. 2024.
\newblock Dl3dv-10k: A large-scale scene dataset for deep learning-based 3d
  vision.
\newblock In \emph{Proceedings of the IEEE/CVF Conference on Computer Vision
  and Pattern Recognition}, 22160--22169.

\bibitem[{Loshchilov and Hutter(2017)}]{loshchilov2017decoupled}
Loshchilov, I.; and Hutter, F. 2017.
\newblock Decoupled weight decay regularization.
\newblock \emph{arXiv preprint arXiv:1711.05101}.

\bibitem[{M{\"u}ller et~al.(2025)M{\"u}ller, Landsgesell, Van~Holland, Stotko,
  and Klein}]{muller2025moment}
M{\"u}ller, J.~U.; Landsgesell, R.~T.; Van~Holland, L.; Stotko, P.; and Klein,
  R. 2025.
\newblock Moment-Based 3D Gaussian Splatting: Resolving Volumetric Occlusion
  with Order-Independent Transmittance.
\newblock \emph{arXiv preprint arXiv:2512.11800}.

\bibitem[{Niedermayr, Stumpfegger, and
  Westermann(2024)}]{niedermayr2024compressed}
Niedermayr, S.; Stumpfegger, J.; and Westermann, R. 2024.
\newblock Compressed 3d gaussian splatting for accelerated novel view
  synthesis.
\newblock In \emph{Proceedings of the IEEE/CVF Conference on Computer Vision
  and Pattern Recognition}, 10349--10358.

\bibitem[{{NVIDIA}(2026)}]{tensorrt}
{NVIDIA}. 2026.
\newblock TensorRT.

\bibitem[{Rakhimov et~al.(2022)Rakhimov, Ardelean, Lempitsky, and
  Burnaev}]{rakhimov2022npbg++}
Rakhimov, R.; Ardelean, A.-T.; Lempitsky, V.; and Burnaev, E. 2022.
\newblock Npbg++: Accelerating neural point-based graphics.
\newblock In \emph{Proceedings of the IEEE/CVF conference on computer vision
  and pattern recognition}, 15969--15979.

\bibitem[{Rijsdijk et~al.(2026)Rijsdijk, Peters, Weinmann, and
  Marroquim}]{rijsdijk2026gaussian}
Rijsdijk, J.; Peters, C.; Weinmann, M.; and Marroquim, R. 2026.
\newblock Gaussian Point Splatting.
\newblock \emph{ACM Transactions on Graphics (TOG)}, 45(4): 1--11.

\bibitem[{Ronneberger, Fischer, and Brox(2015)}]{ronneberger2015u}
Ronneberger, O.; Fischer, P.; and Brox, T. 2015.
\newblock U-net: Convolutional networks for biomedical image segmentation.
\newblock In \emph{International Conference on Medical image computing and
  computer-assisted intervention}, 234--241. Springer.

\bibitem[{R{\"u}ckert, Franke, and Stamminger(2022)}]{ruckert2022adop}
R{\"u}ckert, D.; Franke, L.; and Stamminger, M. 2022.
\newblock Adop: Approximate differentiable one-pixel point rendering.
\newblock \emph{ACM Transactions on Graphics (ToG)}, 41(4): 1--14.

\bibitem[{Scherzer, Yang, and Mattausch(2010)}]{scherzer2010exploiting}
Scherzer, D.; Yang, L.; and Mattausch, O. 2010.
\newblock Exploiting temporal coherence in real-time rendering.
\newblock In \emph{ACM SIGGRAPH ASIA 2010 Courses}, 1--26.

\bibitem[{Schied et~al.(2017)Schied, Kaplanyan, Wyman, Patney, Chaitanya,
  Burgess, Liu, Dachsbacher, Lefohn, and Salvi}]{schied2017spatiotemporal}
Schied, C.; Kaplanyan, A.; Wyman, C.; Patney, A.; Chaitanya, C. R.~A.; Burgess,
  J.; Liu, S.; Dachsbacher, C.; Lefohn, A.; and Salvi, M. 2017.
\newblock Spatiotemporal variance-guided filtering: real-time reconstruction
  for path-traced global illumination.
\newblock In \emph{Proceedings of high performance graphics}, 1--12.

\bibitem[{Thibault.(2023)}]{anthony2023splatapult}
Thibault., A.~J. 2023.
\newblock Splatapult.

\bibitem[{Thies, Zollh{\"o}fer, and Nie{\ss}ner(2019)}]{thies2019deferred}
Thies, J.; Zollh{\"o}fer, M.; and Nie{\ss}ner, M. 2019.
\newblock Deferred neural rendering: Image synthesis using neural textures.
\newblock \emph{Acm Transactions on Graphics (TOG)}, 38(4): 1--12.

\bibitem[{Vogels et~al.(2018)Vogels, Rousselle, McWilliams, R{\"o}thlin,
  Harvill, Adler, Meyer, and Nov{\'a}k}]{vogels2018denoising}
Vogels, T.; Rousselle, F.; McWilliams, B.; R{\"o}thlin, G.; Harvill, A.; Adler,
  D.; Meyer, M.; and Nov{\'a}k, J. 2018.
\newblock Denoising with kernel prediction and asymmetric loss functions.
\newblock \emph{ACM Transactions on Graphics (TOG)}, 37(4): 1--15.

\bibitem[{Yang, Liu, and Salvi(2020)}]{yang2020survey}
Yang, L.; Liu, S.; and Salvi, M. 2020.
\newblock A survey of temporal antialiasing techniques.
\newblock In \emph{Computer graphics forum}, volume~39, 607--621. Wiley Online
  Library.

\bibitem[{Yang et~al.(2025)Yang, Xu, Jiang, Lin, and Dai}]{yang2025virtualized}
Yang, X.; Xu, L.; Jiang, L.; Lin, D.; and Dai, B. 2025.
\newblock Virtualized 3D Gaussians: Flexible Cluster-based Level-of-Detail
  System for Real-Time Rendering of Composed Scenes.
\newblock In \emph{Proceedings of the Special Interest Group on Computer
  Graphics and Interactive Techniques Conference Conference Papers}, 1--11.

\bibitem[{Ye, Wu, and Zhou(2026)}]{ye2026depth}
Ye, K.; Wu, H.; and Zhou, K. 2026.
\newblock Depth Peeling for High-Fidelity Gaussian-Enhanced Surfel Rendering.
\newblock In \emph{Proceedings of the IEEE/CVF conference on computer vision
  and pattern recognition}.

\bibitem[{Zamir et~al.(2021)Zamir, Arora, Khan, Hayat, Khan, Yang, and
  Shao}]{zamir2021multi}
Zamir, S.~W.; Arora, A.; Khan, S.; Hayat, M.; Khan, F.~S.; Yang, M.-H.; and
  Shao, L. 2021.
\newblock Multi-stage progressive image restoration.
\newblock In \emph{Proceedings of the IEEE/CVF conference on computer vision
  and pattern recognition}, 14821--14831.

\bibitem[{Zhu et~al.(2023)Zhu, Zhang, R{\"o}thlin, Papas, and
  Meyer}]{zhu2023denoising}
Zhu, S.; Zhang, X.; R{\"o}thlin, G.; Papas, M.; and Meyer, M. 2023.
\newblock Denoising Production Volumetric Rendering.
\newblock In \emph{ACM SIGGRAPH 2023 Talks}. New York, NY, USA: Association for
  Computing Machinery.

\bibitem[{Zwicker et~al.(2015)Zwicker, Jarosz, Lehtinen, Moon, Ramamoorthi,
  Rousselle, Sen, Soler, and Yoon}]{zwicker2015recent}
Zwicker, M.; Jarosz, W.; Lehtinen, J.; Moon, B.; Ramamoorthi, R.; Rousselle,
  F.; Sen, P.; Soler, C.; and Yoon, S.-E. 2015.
\newblock Recent advances in adaptive sampling and reconstruction for Monte
  Carlo rendering.
\newblock In \emph{Computer graphics forum}, volume~34, 667--681. Wiley Online
  Library.

\end{thebibliography}

\appendix
\setcounter{figure}{0}
\setcounter{table}{0}
\setcounter{equation}{0}
\renewcommand{\thefigure}{A\arabic{figure}}
\renewcommand{\thetable}{A\arabic{table}}
\renewcommand{\theequation}{A\arabic{equation}}
\section{Mobile Baseline and Deployment}
Despite recent years of development, efficient 3D Gaussian Splatting (3DGS) on mobile devices remains an open challenge, primarily bottlenecked by massive Gaussian primitive counts and view-dependent sorting time. 
We evaluate mobile ports of SortFreeGS~\cite{hou2025sort} and MobileGS~\cite{du2026mobile}, with reproduction details provided below.
To overcome these computational bottlenecks, we propose a GPU-NPU parallel framework. 
We evaluate the implementation on a Vivo device equipped with a MediaTek NPU and report throughput for the tested methods.
\tabref{tab:supp_mobile} shows the quantitative result of Mobile 540p throughput (FPS) on vivo x300pro.

\subsection{MobileGS Reproduction}
We faithfully reproduced the optimizations in the released Mobile-GS source code \cite{xiaobiaodu2026mobilegs} , including its view-dependent MLP and sort-free rasterizer. Unlike original 3DGS, Mobile-GS avoids depth sorting by constructing unsorted per-tile Gaussian lists and using order-independent weighted color and transmittance accumulation; we preserve its preprocessing, culling, weighting, and compositing equations. 

For mobile deployment, we additionally execute the full MLP in FP16 on the MediaTek NPU and the rasterizer in Vulkan compute on the GPU, using zero-copy buffers and overlapping NPU inference for frame \(N+1\) with GPU rendering of frame \(N\). We introduce no temporal caching, skipped MLP evaluations, or internal resolution scaling.

\subsection{SortFreeGS Reproduction}
Because SortFreeGS does not release a mobile implementation, we port its released CUDA rasterizer~\cite{yukeli2025sortfreegs} to OpenGL ES 3.1 using a conventional 3DGS-GL pipeline~\cite{anthony2023splatapult}. We faithfully preserve its key changes to 3DGS: learned visibility and depth weighting, removal of depth sorting, weighted-sum compositing, and signed background normalization, together with the reference SH, projection, covariance, raster bound, and alpha-threshold ordering. 

For mobile deployment, we additionally apply algorithm-preserving mobile optimizations, including GPU-resident culling and indirect drawing, geometric rejection before SH evaluation, visibility evaluation before RGB, compact 48-byte projected records, and four-vertex quad expansion. Only SH storage uses FP16; all projection, rasterization, and accumulation remain FP32.

\subsection{Future Work}
We also test our pipeline on iOS devices using Metal shaders. On an iPhone 13 Pro, rasterization and neural reconstruction overlap, resulting in approximately 30 FPS at 720p. On an iPhone 17 Pro, NPU inference takes approximately one-third as long, while GPU rasterization time remains relatively stable. This suggests an opportunity to use a larger reconstruction network on devices with faster neural processing units. Extending the pipeline to other mobile devices, integrated graphics processors, and discrete graphics cards requires platform-specific implementation and evaluation.

\begin{table}[!ht]
\centering
\begin{tabular}{lcccc}
\hline
Method & Truck & Room & DrJohnson & Avg. \\
\hline
3DGS-GL                & 5.46 & 12.8 & 3.2 & 7.15 \\
SS             & 24.8 & 65.8 & \underline{69.4} & 53.3 \\
GPS            & 41.3 & 52.8 & 53.0 & 49.03 \\
SortfreeGS & 32.94 & 22.03 & 31.95 & 28.97 \\
MobileGS & -- & 26.98 & 25.61 & 26.29 \\
Ours (1-spp)                  & \textbf{78.8} & \textbf{112.7} & \textbf{86.3} & \textbf{92.6} \\

Ours (1-spp + S)         & \underline{64.6} & \underline{86.7} & 68.3 & \underline{73.2} \\
\hline
\end{tabular}
\caption{Mobile 540p throughput (FPS) on vivo x300pro with MediaTek tianji9500, using one scene per benchmark. Our renderer uses Vulkan, and \textit{Ours-S} runs in FP16 on the NPU. Avg. is the arithmetic mean over available scene results. The best and second-best values are shown in bold and underlined, respectively.}
\label{tab:supp_mobile}
\end{table}

\section{Stochastic Gaussian Stippling}

\begin{figure}[t!]
    \centering
    \includegraphics[width=\linewidth]{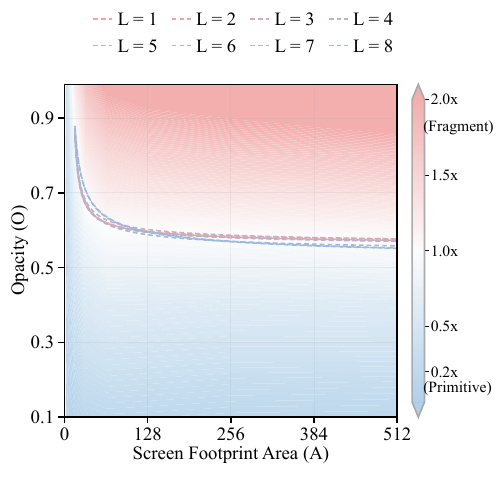}
    \caption{Crossover boundaries ($t_{\mathrm{primitive}} = t_{\mathrm{fragment}}$) for varying layer counts $L \in \{1, \dots, 8\}$ plotted over the opacity-area plane. The almost perfectly overlapping curves indicate that depth complexity has a negligible effect on the relative performance of the two streams.}
    \label{fig:L1-L8}
\end{figure}

For completeness, we summarize the stochastic sampling used by StochasticSplats~\cite{kheradmand2025stochasticsplats} serving as our fragment-based stream, and the opacity-corrected Poisson point process introduced by Gaussian Point Splatting~\cite{rijsdijk2026gaussian} serving as our primitive-based stream. Routing preserves the ordered-compositing expectation under independent Bernoulli acceptance only when each primitive is assigned to one stream and both streams use the same effective alpha function and support. The continuous point construction below uses a pixel-center approximation; it does not by itself establish exact equivalence after discretization or unequal truncation.

Consider a pixel sample location \(\mathbf{x} \in \mathbb{R}^2\) and a projected 2D Gaussian \(\mathcal{G}\) with base opacity \(o\), mean \(\boldsymbol{\mu}\), covariance \(\boldsymbol{\Sigma}\), and
\begin{equation}
\begin{aligned}
G(\mathbf{x}) &= \exp\!\left(-\frac{1}{2}(\mathbf{x} - \boldsymbol{\mu})^\top
\boldsymbol{\Sigma}^{-1}(\mathbf{x} - \boldsymbol{\mu})\right),\\
\alpha(\mathbf{x})& = o G(\mathbf{x}).
\end{aligned}
\end{equation}

The desired probability of selecting primitive \(i\) is

\begin{equation}
\Pr \left[\hat{i}(\mathbf{x}) = i\right] = \alpha_i(\mathbf{x})T_i(\mathbf{x}),
\end{equation}
where \(\hat{i}(\mathbf{x})\) is the index of the closest accepted primitive. This probability follows when the acceptance indicators are independent across primitives and each random variable \(\hat{\alpha}(\mathbf{x})\) satisfies

\begin{equation}
\hat{\alpha}(\mathbf{x})
\sim
\operatorname{Bernoulli}\!\left(\alpha(\mathbf{x})\right),
\end{equation}
With a common effective alpha function and support, this construction gives an unbiased stochastic estimate of the corresponding ordered alpha-compositing result. It does not imply unbiasedness of the subsequent learned reconstruction.

For the fragment-based stream proposed in StochasticSplats, each fragment is generated by a Gaussian primitive following the vanilla 3D Gaussian splatting process. The stochasticity is introduced during the fragment shading stage. Specifically, for a fragment located at pixel \(\mathbf{x}\), a binary random variable \(\hat{\alpha}(\mathbf{x})\) indicates whether the fragment is retained, where the fragment is discarded with probability \(1 - \alpha(\mathbf{x})\). By this construction, \(\hat{\alpha}(\mathbf{x}) \sim \operatorname{Bernoulli}(\alpha(\mathbf{x}))\).

For the primitive-based stream proposed in Gaussian Point Splatting, we first describe a property of a finite-intensity inhomogeneous Poisson point process.
Suppose that the total number of points is \(N \sim \operatorname{Poisson}(\lambda)\). Conditional on \(N\), the point locations \(\Phi=\{X_i\}_{i=1}^{N}\) are independently and identically distributed according to a normalized density \(p\) on \(\mathbb{R}^2\). The resulting point process is an inhomogeneous Poisson point process. For any bounded measurable pixel region \(B \subset \mathbb{R}^2\), we have

\begin{equation}
    N(B)=\#(\Phi\cap B) \sim \operatorname{Poisson}\left(\lambda\int_Bp(\mathbf{y})\;\mathrm{d}\mathbf{y}\right).
\end{equation}
Approximating the density within \(B\) by its value at the pixel center \(\mathbf{x}\), we obtain

\begin{equation}
    \begin{aligned}
    \Pr(N(B) = 0)
    &= \exp\!\left(-\lambda\int_Bp(\mathbf{y})\;\mathrm{d}\mathbf{y}\right)\\
    &\approx e^{-\lambda \lvert B \rvert p(\mathbf{x})}.
    \end{aligned}
\end{equation}
For unit-area pixels, \(\lvert B \rvert = 1\), matching the approximate no-hit probability to \(1-\alpha(\mathbf{x})\) gives

\begin{equation}
    p(\mathbf{x}) = -\frac{1}{\lambda}\ln(1 - \alpha(\mathbf{x})).
\end{equation}

Following the Box--Muller construction, let \(LL^\top = \boldsymbol{\Sigma}\) and \(\mathbf{x}(r,\theta)=\boldsymbol{\mu}+L(r\cos\theta,r\sin\theta)^\top\). Including the change-of-variables Jacobian gives the polar-coordinate density

\begin{equation}
    \begin{aligned}
    p_{r,\theta}(r,\theta)
    &= \lvert\det L\rvert r\,p(\mathbf{x}(r,\theta))\\
    &= -\frac{\lvert\det L\rvert}{\lambda}r\ln\left(1-o e^{-\frac{r^2}{2}}\right).
    \end{aligned}
\end{equation}

The angle is uniform on \([0,2\pi)\), and the cumulative distribution function (CDF) of \(r\) is

\begin{equation}
\begin{aligned}
    F(r) &= 2\pi \int_0^r -\frac{\lvert\det L\rvert}{\lambda}\ln\left(1-o e^{-\frac{u^2}{2}}\right)u\;\mathrm{d}u \\
    &= \frac{2\pi\lvert\det L\rvert}{\lambda}\left(\operatorname{Li}_2(o)-\operatorname{Li}_2\left(o e^{-\frac{r^2}{2}}\right)\right).
\end{aligned}
\end{equation}
Since \(F(\infty) = 1\), solving for \(\lambda\) gives

\begin{equation}
    \lambda  = 2\pi \sqrt{\lvert \boldsymbol{\Sigma} \rvert} \operatorname{Li}_2(o),
\end{equation}
and therefore 
\begin{equation}
    F(r) = 1 - \operatorname{Li}_2\left(o e^{-\frac{r^2}{2}}\right) / \operatorname{Li}_2(o).
\end{equation}
Sampling an independent random variable
\(u_1 \sim \operatorname{Uniform}(0, 1)\) and applying inverse transform sampling with \(F(r) = u_1\), we have

\begin{equation}
    r = \sqrt{-2\ln\frac{\operatorname{Li}_2^{-1}((1 - u_1) \operatorname{Li}_2(o))}{o}}.
\end{equation}
Because the sampling function for \(r\) differs from that of the original Box--Muller transform, we refer to this procedure as the opacity-corrected Box--Muller transform.

In summary, sampling the total number of points as
\(N \sim \operatorname{Poisson}(\lambda)\), followed by opacity-corrected Box--Muller sampling for each point location, constructs an inhomogeneous Poisson point process. Its pixel-occupancy indicator is Bernoulli with exact success probability \(1-\exp(-\lambda\int_Bp(\mathbf{y})\,\mathrm{d}\mathbf{y})\), which approximates \(\alpha(\mathbf{x})\) under the unit-area pixel-center approximation above. Finite-support truncation must be handled consistently across streams before asserting the same ordered-compositing expectation.

\begin{table*}[tbhb]
\centering
\begin{tabular}{lcccccc}
\hline
\multirow{2}{*}{Scene}
& \multicolumn{3}{c}{Small (1 spp)}
& \multicolumn{3}{c}{Large (1 spp)} \\
\cline{2-4}\cline{5-7}
& PSNR$\uparrow$ & SSIM$\uparrow$ & LPIPS$\downarrow$ 
& PSNR$\uparrow$ & SSIM$\uparrow$ & LPIPS$\downarrow$ \\
\hline
Bicycle   & 24.0400 & 0.6710 & 0.2600 & 24.6107 & 0.6996 & 0.2745 \\
Bonsai    & 30.7500 & 0.9110 & 0.1600 & 32.1253 & 0.9284 & 0.2303 \\
Counter   & 28.2700 & 0.8690 & 0.1890 & 28.9801 & 0.8927 & 0.2317 \\
Garden    & 25.4000 & 0.7470 & 0.1750 & 26.0137 & 0.7769 & 0.1891 \\
Kitchen   & 29.2100 & 0.8750 & 0.1550 & 30.0836 & 0.8948 & 0.1703 \\
Room      & 30.6000 & 0.8870 & 0.2090 & 31.3050 & 0.9048 & 0.2527 \\
Stump     & 25.7900 & 0.7160 & 0.2160 & 26.1589 & 0.7371 & 0.2509 \\
\textbf{Mip-NeRF 360 Avg.} 
          & \textbf{27.7229} & 0.8120 & 0.1949 
          & \textbf{28.4682} & \textbf{0.8335} & \textbf{0.2285} \\
\hline
Train     & 21.4500 & 0.7400 & 0.2360 & 22.1772 & 0.7800 & 0.2476 \\
Truck     & 23.9900 & 0.8040 & 0.1790 & 24.7056 & 0.8376 & 0.2081 \\
\textbf{Tanks \& Temples Avg.} 
          & \textbf{22.7200} & 0.7780 & 0.2075 
          & \textbf{23.4414} & \textbf{0.8088} & \textbf{0.2279} \\
\hline
DrJohnson & 28.8000 & 0.8820 & 0.1980 & 29.1139 & 0.9010 & 0.2565 \\
Playroom  & 29.8600 & 0.8890 & 0.1810 & 30.1277 & 0.9085 & 0.2550 \\
\textbf{Deep Blending Avg.} 
          & \textbf{29.3300} & 0.8952 & 0.1895 
          & \textbf{29.6208} & \textbf{0.9048} & \textbf{0.2558} \\
\hline
\textbf{All Scenes Avg.} 
          & \textbf{27.1055} & \textbf{0.8174} & \textbf{0.1962} 
          & \textbf{27.7638} & \textbf{0.8419} & \textbf{0.2333} \\
\hline
\end{tabular}
\caption{Per-scene and per-dataset reconstruction quality of the Small and Large variants at 1 spp, measured against ground-truth images. Dataset averages are arithmetic means over their constituent scenes; the final row is the macro-average over all 11 scenes.}
\label{tab:1spp}
\end{table*}

\begin{table*}[t!]
\centering
\begin{tabular}{lcccccc}
\hline
\multirow{2}{*}{Scene}
& \multicolumn{3}{c}{Small (4 spp)}
& \multicolumn{3}{c}{Large (4 spp)} \\
\cline{2-4}\cline{5-7}
& PSNR$\uparrow$ & SSIM$\uparrow$ & LPIPS$\downarrow$
& PSNR$\uparrow$ & SSIM$\uparrow$ & LPIPS$\downarrow$ \\
\hline
Bicycle  & 24.5220 & 0.7018 & 0.2799 & 24.9213 & 0.7212 & 0.2614 \\
Bonsai   & 31.2651 & 0.9232 & 0.2346 & 32.4673 & 0.9317 & 0.2279 \\
Counter  & 28.5755 & 0.8876 & 0.2357 & 29.1298 & 0.8981 & 0.2263 \\
Garden   & 26.1494 & 0.7916 & 0.1903 & 26.5915 & 0.8104 & 0.1662 \\
Kitchen  & 29.9029 & 0.8948 & 0.1701 & 30.5050 & 0.9057 & 0.1595 \\
Room     & 30.9545 & 0.9004 & 0.2581 & 31.5079 & 0.9085 & 0.2485 \\
Stump    & 26.2412 & 0.7415 & 0.2563 & 26.4615 & 0.7535 & 0.2404 \\
\textbf{Mip-NeRF 360 Avg.}
         & \textbf{28.2301} & \textbf{0.8344} & \textbf{0.2321}
         & \textbf{28.7978} & \textbf{0.8470} & \textbf{0.2186} \\
\hline
Train    & 21.5588 & 0.7535 & 0.2913 & 22.1426 & 0.7808 & 0.2489 \\
Truck    & 24.3956 & 0.8274 & 0.2398 & 25.0175 & 0.8451 & 0.2083 \\
\textbf{Tanks \& Temples Avg.}
         & \textbf{22.9772} & \textbf{0.7904} & \textbf{0.2656}
         & \textbf{23.5800} & \textbf{0.8130} & \textbf{0.2286} \\
\hline
DrJohnson & 28.9455 & 0.8978 & 0.2654 & 29.1475 & 0.9029 & 0.2562 \\
Playroom  & 30.0227 & 0.9019 & 0.2528 & 30.2009 & 0.9086 & 0.2529 \\
\textbf{Deep Blending Avg.}
          & \textbf{29.4841} & \textbf{0.8999} & \textbf{0.2591}
          & \textbf{29.6742} & \textbf{0.9058} & \textbf{0.2545} \\
\hline
\textbf{All Scenes Avg.}
          & \textbf{27.5030} & \textbf{0.8383} & \textbf{0.2431}
          & \textbf{28.0084} & \textbf{0.8515} & \textbf{0.2270} \\
\hline
\end{tabular}
\caption{Per-scene and per-dataset reconstruction quality of the Small and Large variants at 4 spp, measured against ground-truth images. }
\label{tab:4spp}
\end{table*}

\begin{table*}[t!]
\centering
\begin{tabular}{lcccccc}
\hline
\multirow{2}{*}{Scene}
& \multicolumn{3}{c}{Small (16 spp)}
& \multicolumn{3}{c}{Large (16 spp)} \\
\cline{2-4}\cline{5-7}
& PSNR$\uparrow$ & SSIM$\uparrow$ & LPIPS$\downarrow$ 
& PSNR$\uparrow$ & SSIM$\uparrow$ & LPIPS$\downarrow$ \\
\hline
Bicycle   & 24.9336 & 0.7303 & 0.2550 & 25.1800 & 0.7470 & 0.2080 \\
Bonsai    & 31.7702 & 0.9304 & 0.2228 & 32.8700 & 0.9350 & 0.1430 \\
Counter   & 28.7965 & 0.8974 & 0.2180 & 29.2900 & 0.9000 & 0.1690 \\
Garden    & 26.7682 & 0.8296 & 0.1487 & 27.0900 & 0.8450 & 0.1000 \\
Kitchen   & 30.4227 & 0.9098 & 0.1461 & 30.7800 & 0.9150 & 0.1120 \\
Room      & 31.2549 & 0.9094 & 0.2407 & 31.7000 & 0.9110 & 0.1850 \\
Stump     & 26.4946 & 0.7595 & 0.2347 & 26.6500 & 0.7660 & 0.1760 \\
\textbf{Mip-NeRF 360 Avg.} 
          & \textbf{28.6344} & \textbf{0.8523} & \textbf{0.2094} 
          & \textbf{29.0757} & 0.8606 & 0.1561 \\
\hline
Train     & 21.7615 & 0.7754 & 0.2633 & 22.1800 & 0.7880 & 0.1910 \\
Truck     & 24.7443 & 0.8469 & 0.2036 & 25.1800 & 0.8560 & 0.1230 \\
\textbf{Tanks \& Temples Avg.} 
          & \textbf{23.2529} & \textbf{0.8112} & \textbf{0.2335} 
          & \textbf{23.6800} & 0.8252 & 0.1570 \\
\hline
DrJohnson & 29.0140 & 0.8998 & 0.2544 & 29.1600 & 0.8940 & 0.1810 \\
Playroom  & 30.1152 & 0.9062 & 0.2419 & 30.2400 & 0.9030 & 0.1660 \\
\textbf{Deep Blending Avg.} 
          & \textbf{29.5646} & \textbf{0.9030} & \textbf{0.2482} 
          & \textbf{29.7000} & 0.9065 & 0.1735 \\
\hline
\textbf{All Scenes Avg.} 
          & \textbf{27.8251} & \textbf{0.8541} & \textbf{0.2208} 
          & \textbf{28.2109} & \textbf{0.8605} & \textbf{0.1595} \\
\hline
\end{tabular}
\caption{Per-scene and per-dataset reconstruction quality of the Small and Large variants at 16 spp, measured against ground-truth images. Dataset averages are arithmetic means over their constituent scenes; the final row is the macro-average over all 11 scenes.}
\label{tab:16spp}
\end{table*}

\begin{table}[t!]
\centering


\begin{tabular}{lcccc}
\hline
Variant & Bonsai & Counter & Train & Avg. \\
\hline
Forward Nearest.  & \textbf{30.75} & \underline{28.23} & \textbf{21.50} & \textbf{26.83} \\
Forward Bilinear. & 30.60 & 28.19 & 21.38 & 26.72 \\
Backward Nearest. & 30.59 & 28.21 & \underline{21.40} & 26.73 \\
Backward Bilinear.& \underline{30.68} & \textbf{28.25} & \underline{21.40} & \underline{26.78} \\
\hline
\end{tabular}



\caption{Temporal transport under stochastic visibility. Backward gathering produces structured artifacts; bilinear transport spreads samples, whereas nearest-forward transport preserves discrete observations. The best and second-best PSNR (dB) values are shown in \textbf{bold} and \underline{underlined}; ties share the same rank.}
\label{tab:ablation_temporal}
\end{table}

\section{Route Criterion Estimation}
As discussed in the main text, we evaluate the relative time cost of the primitive-based and fragment-based streams under varying layer counts $L \in \{1, \dots, 8\}$ to understand the impact of inter-layer occlusion. 

In \figref{fig:L1-L8}, we visualize the crossover boundaries—the zero-ratio contours where $t_{\mathrm{primitive}} = t_{\mathrm{fragment}}$ (or $S(o, A) = 0$)—for $L$ ranging from 1 to 8 on a single opacity-area coordinate system. 
As observed, these eight boundary curves almost overlap. 
This demonstrates that depth complexity and occlusion relationships have a minimal impact on the relative execution time of the two rendering streams. 
Consequently, our routing criterion is robust to varying occlusion conditions, justifying the use of a unified routing rule regardless of the scene's depth complexity.

\section{Implement Details}

\subsection{Dataset}
We use custom scene names for clarity when presenting the DL3DV~\cite{ling2024dl3dv} scenes. The corresponding hash identifiers in the dataset are listed in \tabref{tab:scene_hashes}.

\begin{table*}[t]
\centering
\begin{tabular}{l l}
\toprule
\textbf{Scene} & \textbf{Hash Value} \\
\midrule
Pavil. & \texttt{d3812aad538261e7f73c75762ff55f23b468bcc76f376d52ac86ca6cf3c44b4b} \\
Stat.   & \texttt{c37109a55effe0000f8e40652ca935376e75bcb2a0b56de8eabd20a26e2a0f68} \\
Flow.  & \texttt{1d6a9ed47cce39fd1c4d18f776bcc97e507b81bde921ca596bd91b0b02b5e414} \\
Plant    & \texttt{dafa9c7cbda9d1ddaa8a2b51fc8c54f4eb44161f5e5c53685dc744580ca77751} \\
House    & \texttt{6d81c5ab0d480fd43d78b75ff372a8113ad38e2c03f1d69627c009883054d4c2} \\
\bottomrule

\end{tabular}

\caption{Scene names and corresponding dataset hash values in DL3DV.}
\label{tab:scene_hashes}
\end{table*}

\subsection{Compare with Gaussian Point Splatting}
\begin{figure}[t!]
    \centering
    \includegraphics[width=\linewidth]{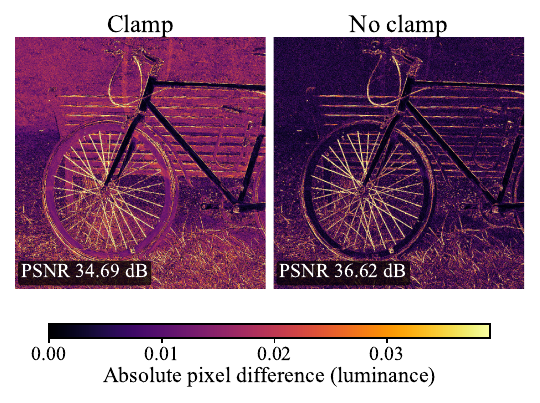}
    \caption{After 1024spp rendering using GPS's source code, the result was compared with that of Gaussian rendering, and there was a loss. The only modification to Gaussian Point Splatting is that we removed the upper limit on individual Gaussian sampling.}
    \label{fig:clamp_bias}
\end{figure}

\begin{figure}[t]
    \centering
    \includegraphics[width=\linewidth]{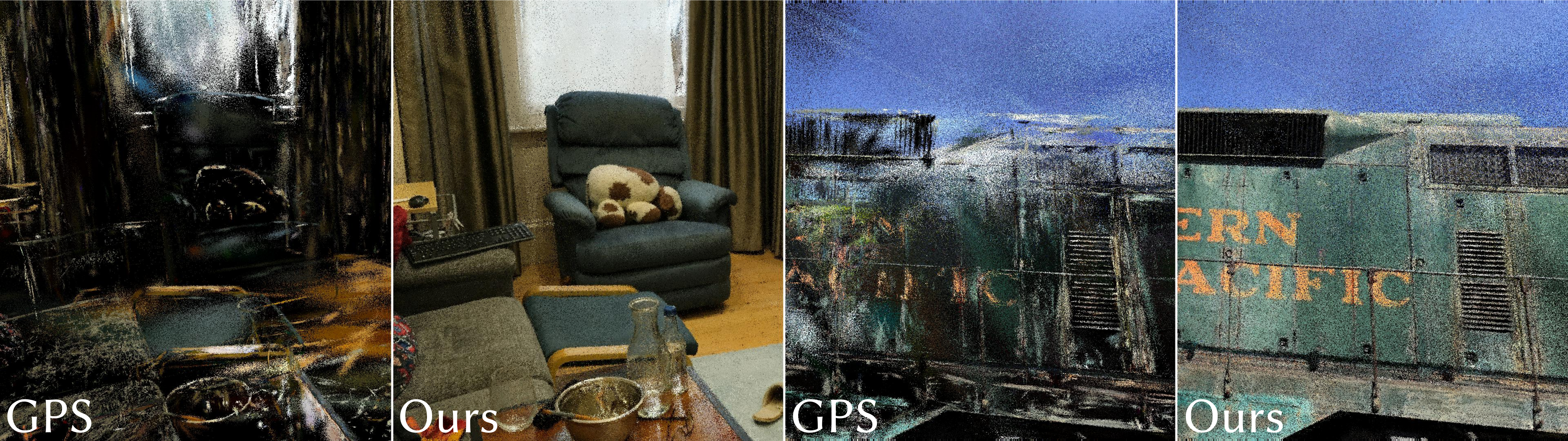}
    \caption{Rendering failure of Gaussian Point Splatting: its workload distribution loses much detail when aligning the 3DGS bounding box at conventional rendering resolution.}
    \label{fig:failurecase}
\end{figure}

Our runtime comparison with GPS strictly follows their original implementation, with one exception: we disabled its upper limit on single-Gaussian sampling. 
As illustrated in the ~\figref{fig:clamp_bias}, artificially capping this limit leads to severe artifacts from certain viewpoints. 
As a concurrent work, the only architectural difference between our Fragment Stream and GPS lies in the retrieval of Gaussian IDs. 
Specifically, we launch $N_p$ threads and perform a binary search over a prefix-sum array to locate the corresponding Gaussian IDs, whereas they utilize a pre-allocated fixed array. 
Theoretically, this binary search introduces global memory queries with a time complexity of $\mathcal{O}(N_{\text{sample}} \log N_{\text{Gaussian}})$. 
In practice, however, the observed runtime difference is negligible. 
We attribute this to the fact that the rendering pipeline is primarily compute-bound rather than memory I/O-bound. 
Furthermore, GPS imposes a hard hyperparameter limit of $2.5 \times 10^8$ total sampling points, which triggers forced culling and ultimately results in rendering failures, as depicted in \figref{fig:failurecase}.
While increasing this hyperparameter can alleviate the issue, we typically lack prior knowledge of novel scene characteristics and arbitrary user viewpoints. Consequently, it is intractable to determine an optimal amount of VRAM for pre-allocation (e.g., a cap of $2.5 \times 10^8$ sampling points consumes approximately 1 GB of memory).

\subsection{Network Architecture}
\paragraph{Network architecture.}
We propose two variants \textit{Ours-S} and \textit{Ours-L} to test the limit of the reconstructor.
Both variants take the current frame and \(K = 3\) reprojected history frames as input. Each pixel \(\mathbf{x}\) contributes 10 channels:

\begin{itemize}
    \item RGB color \(\mathbf{c}(\mathbf{x})\).
    \item alpha \(\alpha(\mathbf{x})\).
    \item squared Mahalanobis distance
    \[\delta^2(\mathbf{x})=(\mathbf{x}-\boldsymbol{\mu})^\top
\boldsymbol{\Sigma}^{-1}(\mathbf{x}-\boldsymbol{\mu}).\]
    \item opacity \(o\) of the selected Gaussian primitive.
    \item independent projected 2D covariance coefficients of the selected Gaussian primitive
    \[\mathbf{q}=(\boldsymbol{\Sigma}_{0, 0},\boldsymbol{\Sigma}_{0, 1},\boldsymbol{\Sigma}_{1, 1})^\top.\]
    \item inverse depth \(z^{-1}\) of the selected Gaussian primitive.
\end{itemize}
These concatenated features result in \((K + 1) \times 10 = 40\) input channels.

\begin{displaymath}
\begin{aligned}
F_{t}^{\mathrm{local}}
    =P(X_t),\qquad F_{t}^{\mathrm{context}} =U_{\theta}(X_t),\\
\hat I_t
    =H_{\theta}\!\left(
      \operatorname{concat}\left(
      F_{t}^{\mathrm{local}},
      F_{t}^{\mathrm{context}}
      \right)\right),
\end{aligned}
\end{displaymath}

Both variants use encoder--decoder U-Net architectures without normalization. Hidden convolutions are followed by ReLU activations. Downsampling uses \(2\times2\) max pooling, while each decoder stage uses nearest-neighbor upsampling, concatenation with the corresponding encoder feature, and two \(3\times3\) convolutions.

\textit{Ours-S} uses three downsampling stages with a constant feature width of 16 channels. Its 40-channel input is first projected to 16 channels using a \(1\times1\) convolution. A separate \(1\times1\) convolution projects the full-resolution 40-channel input to a 16-channel skip feature.

\textit{Ours-L} uses four downsampling stages with encoder widths \(64,96,128,192,\) and \(256\). At full resolution, its 40-channel input is projected to a 32-channel skip feature using a \(1\times1\) convolution.

A final \(3\times3\) convolution maps the decoder features to three RGB channels. \textit{Ours-S} uses a linear output, whereas \textit{Ours-L} applies ReLU after the final convolution. Both variants predict RGB directly and do not use residual output addition.

\begin{table*}[t]
\centering
\setlength{\tabcolsep}{5pt}
\begin{tabular}{@{}lcc@{}}
\hline
Configuration & Ours-S & Ours-L \\
\hline
Input / output channels
    & $40/3$ & $40/3$ \\
Down-/upsampling stages
    & $3/3$ & $4/4$ \\
Resolution pyramid
    & $H$ to $H/8$ & $H$ to $H/16$ \\
Encoder widths, fine to coarse
    & $16,16,16,16$
    & $64,96,128,192,256$ \\
Encoder/bottleneck convolutions per resolution
    & \shortstack{$2,1,1,2$\\
      (first stage: $1{\times}1$, then $3{\times}3$)}
    & \shortstack{$2,2,2,2,2$\\
      (all $3{\times}3$)} \\
Decoder transitions, coarse to fine
    & \shortstack{$32\!\to\!16\!\to\!16$ at $H/4$;\\
                  $32\!\to\!16\!\to\!16$ at $H/2$;\\
                  $32\!\to\!16\!\to\!16$ at $H$}
    & \shortstack{$384\!\to\!192\!\to\!192$ at $H/8$;\\
                  $288\!\to\!128\!\to\!128$ at $H/4$;\\
                  $192\!\to\!96\!\to\!96$ at $H/2$;\\
                  $128\!\to\!64\!\to\!64$ at $H$} \\
Convolutions per decoder stage
    & two $3\times3$
    & two $3\times3$ \\
Full-resolution skip projection
    & $40\!\to\!16$ ($1\times1$)
    & $40\!\to\!32$ ($1\times1$) \\
Output head
    & $16\!\to\!3$ ($3\times3$)
    & $64\!\to\!3$ ($3\times3$) \\
Activation / normalization
    & hidden ReLU, linear output / none
    & ReLU including output / none \\
Output formulation
    & direct RGB
    & direct RGB \\
Downsampling / upsampling
    & max pool / nearest
    & max pool / nearest \\
Total convolutional layers
    & $14$ ($12$ conv-$3$ + $2$ conv-$1$)
    & $20$ ($19$ conv-$3$ + $1$ conv-$1$) \\
Parameters
    & $34{,}179$
    & $3{,}880{,}803$ \\
FP16 kernel-weight size
    & $0.0648$ MiB
    & $7.397$ MiB \\
\hline
\end{tabular}
\caption{Detailed reconstruction-network architectures. A decoder
transition $a\!\to\!b\!\to\!c$ denotes the channel count after skip
concatenation and the outputs of the two subsequent $3\times3$
convolutions. Both variants use concatenative encoder skips, a projected
full-resolution input skip, max-pool downsampling, nearest-neighbor
upsampling, and direct RGB prediction. Parameter counts include
convolutional biases. FP16 kernel-weight sizes exclude biases,
activations, and runtime workspace.}
\label{tab:supp_architecture}
\end{table*}

The renderer-supervised generalization model uses the same \textit{Ours-S} architecture and loss as the scene-specific model. 

\tabref{tab:supp_architecture} shows the network details. \tabref{tab:1spp}, \tabref{tab:4spp}, and \tabref{tab:16spp} show reconstruction quality at different sampling budgets and network sizes.

\subsection{Temporal Accumulation}
In our interactive GUI app, we use a simple reprojected temporal accumulation of reconstructed images to further improve the temporal stability.

The accumulated result from the previous frame is forward-reprojected into the current camera, with depth competition used to retain the closest history sample. Valid history is combined with the current denoised output using an incremental average with a capped history length: 

\[
A_t(\mathbf{x})
=
\left(1-\frac{1}{N_t}\right)
\widetilde{A}_{t-1}(\mathbf{x})
+
\frac{1}{N_t}D_t(\mathbf{x}),
\]
where \(D_t\) is the current denoised frame and
\(\widetilde{A}_{t-1}\) is the forward-reprojected accumulated result.
We use \(N_t=12\) for a stationary camera and \(N_t=2\) during camera
motion. Invalid history is replaced by \(D_t\).

\section{Extended Ablations}

\subsection{Temporal Ablations}

We evaluated forward and backward reprojection combined with nearest-neighbor and bilinear sampling in \tabref{tab:ablation_temporal}.
Forward nearest-neighbor reprojection achieves the highest mean PSNR among the evaluated variants.

\begin{figure}[t]
    \centering
    \includegraphics[width=\linewidth]{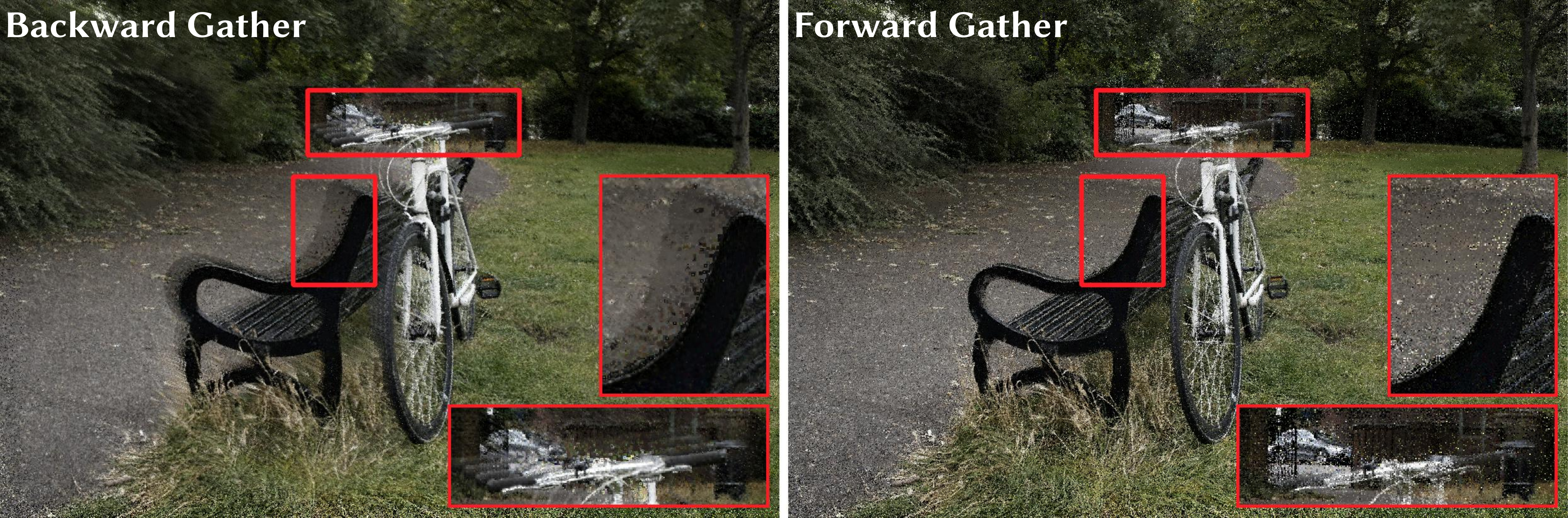}
    \caption{Temporal ablations in the evaluated stochastic-input setting. Backward reprojection exhibits structured artifacts, while forward reprojection leaves unfilled pixels.}
    \label{fig:ghost}
\end{figure}

Traditional temporal information reuse typically relies on either forward or backward reprojection, both of which exhibit inherent limitations. 
Forward reprojection can leave holes where history does not cover the target view. In our stochastic-input setting, backward reprojection can introduce structured artifacts, as illustrated in \figref{fig:ghost}.
With the evaluated network capacity, forward nearest-neighbor reprojection yields higher mean reconstruction PSNR than either backward variant.

Furthermore, regarding the processing of historical samples, we investigated both nearest-neighbor and bilinear interpolation. 
Under forward reprojection, nearest-neighbor sampling yields higher PSNR than bilinear sampling on the three evaluated scenes.
One possible explanation is that nearest-neighbor assignment avoids mixing descriptor values from neighboring observations.

\begin{figure*}[t]
    \centering
    \includegraphics[width=\linewidth]{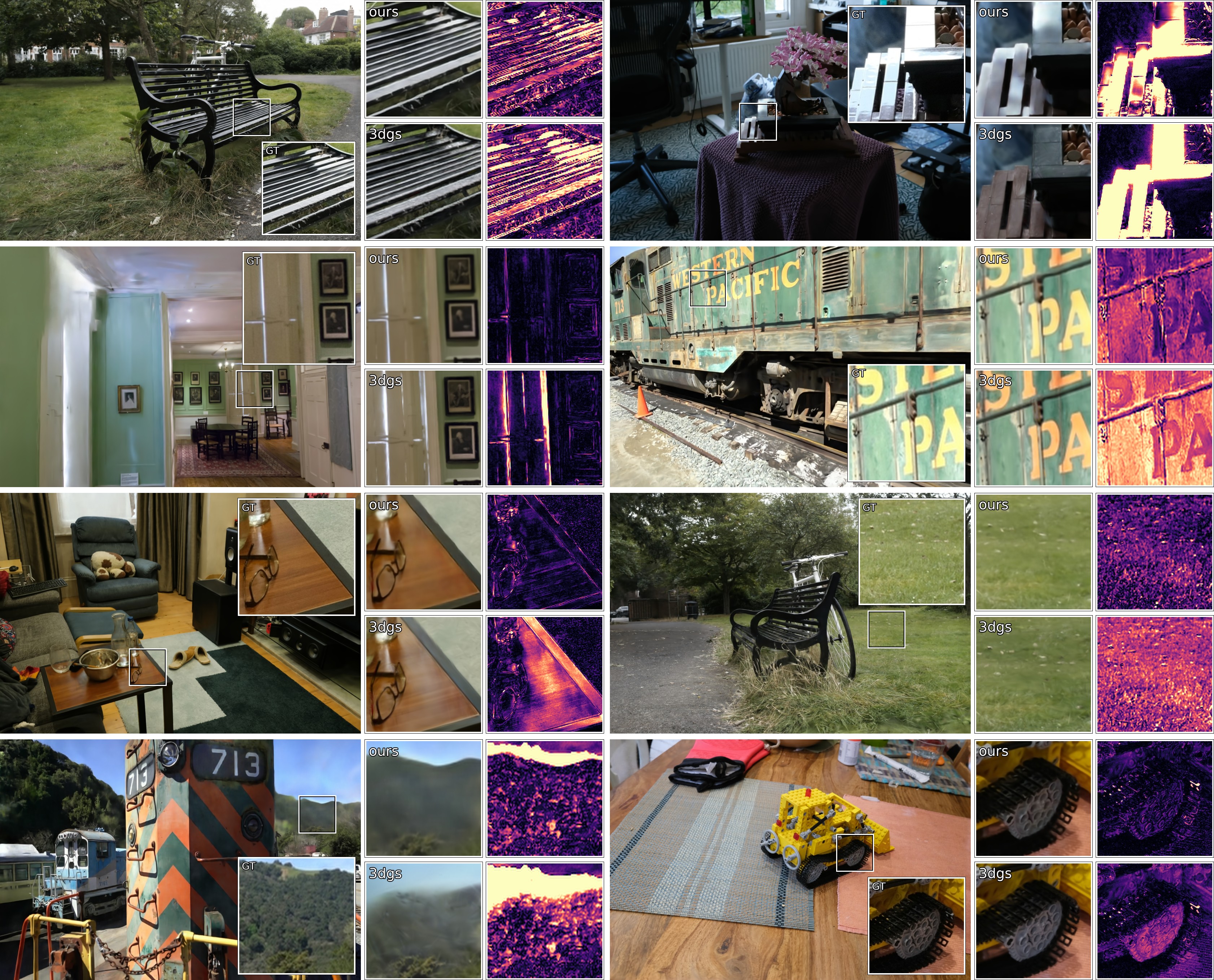}
    \caption{Qualitative results of additional scenes of 16spp + Large network showing neural reconstruction quality}
    \label{fig:additional_qualitative}
\end{figure*}

\subsection{Observation Budget and Network Size Ablations}

We evaluate rendering quality across scenes using sampling budgets of 1, 4, and 16 spp and two network capacities.
The compact network \textit{Ours-S} reduces stochastic noise, while the larger \textit{Ours-L} provides additional capacity for reconstructing local appearance details.
Qualitative examples are provided in the main text and \secref{sec:additional}.

In terms of practical implementation, we align our framework with traditional rendering pipelines by integrating Temporal Anti-Aliasing (TAA) subsequent to the neural denoising module. 
The supplementary video shows this combined pipeline for the 1-spp + \textit{Ours-S} configuration. Its temporal appearance reflects both neural reconstruction and the additional TAA stage.

\section{Additional Qualitative Results}
\label{sec:additional}
~\figref{fig:additional_qualitative} shows large networks can reconstruct complex lighting and local details.

\end{document}